%
\magnification = \magstep1
\hsize = 15.5 truecm
\vsize = 24.0 truecm
\baselineskip  = 12.0pt 
\parindent     = 20.0pt
\overfullrule=0pt
\parskip       =  0.0pt 

\font\top=cmbx12 scaled \magstep1
\font\toph=cmbx10 scaled \magstep2

\def \la {\mathrel{\vcenter
     {\offinterlineskip \hbox{$<$}\hbox{$\sim$}}}}
\def \ga {\mathrel{\vcenter
     {\offinterlineskip \hbox{$>$}\hbox{$\sim$}}}}
\def\simgr{\,\hbox{\hbox{$ > $}\kern -0.8em \lower
     1.0ex\hbox{$\sim$}}\,}
\def\simle{\,\hbox{\hbox{$ < $}\kern -0.8em \lower
     1.0ex\hbox{$\sim$}}\,}
%
%
\def \refs {\begingroup \frenchspacing
            \parskip = 0.2 \baselineskip \parindent = 0 pt
            \everypar = {\hangindent = 20.0 pt \hangafter = 1}}
\def \endrefs {\par \endgroup}
\def\ref#1{\lbrack #1\rbrack}
\def\eck#1{\left\lbrack #1 \right\rbrack}
\def\rund#1{\left( #1 \right)}
\def\ave#1{\langle #1 \rangle}

%
%
\font \tbfontt                = cmbx10 scaled\magstep1
\font \ninebf                 = cmbx9
\font \ninei                  = cmmi9
\font \nineit                 = cmti9
\font \ninerm                 = cmr9
\font \ninesans               = cmss10 at 9pt
\font \ninesl                 = cmsl9
\font \ninesy                 = cmsy9
\font \ninett                 = cmtt9
\font \sixbf                  = cmbx6
\font \sixi                   = cmmi6
\font \sixrm                  = cmr6
\font \fivesans               = cmss10 at 5pt
\font \sixsans                = cmss10 at 6pt
\font \sevensans              = cmss10 at 7pt
\font \ninesans               = cmss10 at 9pt
\font \tensans                = cmss10
\font \sixsy                  = cmsy6
\newfam\sansfam
\textfont\sansfam=\tensans\scriptfont\sansfam=\sevensans
\scriptscriptfont\sansfam=\fivesans
\def\petit{\def\rm{\fam0\ninerm}%
\textfont0=\ninerm \scriptfont0=\sixrm \scriptscriptfont0=\fiverm
 \textfont1=\ninei \scriptfont1=\sixi \scriptscriptfont1=\fivei
 \textfont2=\ninesy \scriptfont2=\sixsy \scriptscriptfont2=\fivesy
 \def\it{\fam\itfam\nineit}%
 \textfont\itfam=\nineit
 \def\sl{\fam\slfam\ninesl}%
 \textfont\slfam=\ninesl
 \def\bf{\fam\bffam\ninebf}%
 \textfont\bffam=\ninebf \scriptfont\bffam=\sixbf
 \scriptscriptfont\bffam=\fivebf
 \def\sans{\fam\sansfam\ninesans}%
 \textfont\sansfam=\ninesans \scriptfont\sansfam=\sixsans
 \scriptscriptfont\sansfam=\fivesans
 \def\tt{\fam\ttfam\ninett}%
 \textfont\ttfam=\ninett
 \normalbaselineskip=11pt
 \setbox\strutbox=\hbox{\vrule height7pt depth2pt width0pt}%
 \normalbaselines\rm}
\def\endpet{\vskip6pt\egroup}
\def\begref{\par\bgroup\petit
\let\sl=\rm\let\INS=N}
\def\ref{\goodbreak\if N\INS\let\INS=Y
\goodbreak\vskip25true pt plus4pt minus4pt
\noindent{\tbfontt References}\vskip15true pt plus4pt minus4pt
\mark{References}\fi
\hangindent0.5\parindent\hangafter=1\noindent\ignorespaces}
\def\begfig#1cm#2\endfig{\par
\setbox1=\vbox{\dimen0=#1true cm\advance\dimen0
by1true cm\kern\dimen0#2}%
\dimen0=\ht1\advance\dimen0by\dp1\advance\dimen0by5\baselineskip
\advance\dimen0by0.4true cm
\ifdim\dimen0>\vsize\pageinsert\box1\vfill\endinsert
\else
\dimen0=\pagetotal\ifdim\dimen0<\pagegoal
\advance\dimen0by\ht1\advance\dimen0by\dp1\advance\dimen0by1.4true cm
\ifdim\dimen0>\vsize
\topinsert\box1\endinsert
\else\box1\vskip4true mm\fi
\else\box1\vskip4true mm\fi\fi}
\def\begdoublefig#1cm #2 #3 \enddoublefig{\begfig#1cm%
\vskip-.8333\baselineskip\line{\vtop{\hsize=0.46\hsize#2}\hfill
\vtop{\hsize=0.46\hsize#3}}\endfig}
\let\ts = \thinspace
\def\figure#1#2{\vskip1true cm\setbox0=\vbox{\noindent\petit{\bf Fig.\ts#1.\
}\ignorespaces #2\smallskip
\count255=0\global\advance\count255by\prevgraf}%
\ifnum\count255>1\box0\else
\centerline{\petit{\bf Fig.\ts#1.\ }\ignorespaces#2}\smallskip\fi}
%
\input epsf
%
%
$\phantom{.}$
\vskip 0.5truecm
\centerline {{\toph Numerical models of}}
\smallskip
\smallskip
\centerline {{\toph protoneutron stars and type-II supernovae ---}}
\smallskip
\smallskip
\centerline {{\toph recent developments}}
%
\bigskip\smallskip
\centerline {{\bf H.-Thomas Janka}}
\medskip
\centerline {Max-Planck-Institut f\"ur Astrophysik}
\centerline {Karl-Schwarzschild-Str.~1, D-85740 Garching, Germany}
\centerline{{\tt email: thj@mpa-garching.mpg.de}}
\bigskip\bigskip\bigskip\noindent
\centerline{ABSTRACT}
\medskip\smallskip\noindent
The results of recent multi-dimensional simulations of type-II
supernovae are reviewed. They show that convective 
instabilities in the collapsed stellar core 
might play an important role already during the
first second after the formation of the supernova shock.
Convectively unstable situations occur below and near
the neutrinosphere as well as in the neutrino-heated
region between the nascent neutron star and the 
supernova shock after the latter has stalled at a radius of
typically 100--200~km. 

While convective overturn in the
layer of neutrino energy deposition clearly helps the 
explosion to develop and potentially provides an explanation of
strong mantle and envelope mixing, asphericities, and 
non-uniform $^{56}$Ni distribution observed in supernova SN~1987A,
its presence and importance depends on the strength of 
the neutrino heating and thus on the size of the neutrino 
fluxes from the neutron star. Convection in the hot-bubble region 
can only develop if the growth timescale of the instabilities 
and the heating timescale are both shorter than the accretion
timescale of the matter advected through the stagnant shock. For
too small neutrino luminosities this requirement is not fulfilled
and convective activity cannot develop, leading to very weak 
explosions or even fizzling models, just as in the one-dimensional
situation. 

Convectively enhanced neutrino luminosities from the protoneutron 
star can therefore provide an essential condition for the explosion
of the star. Very recent two-dimensional, self-consistent,
general relativistic simulations of the cooling of a newly-formed 
neutron star demonstrate and 
confirm the possibility that Ledoux convection, driven by negative
lepton number and entropy gradients, may encompass the whole
protoneutron star within less than one second and can lead
to an increase of the neutrino fluxes by up to a factor of two.
\smallskip\bigskip\bigskip\noindent
{\bf 1.$\ $Introduction}
\medskip\smallskip\noindent
Neutrinos play a crucial role in our understanding of type-II
supernova explosions. According to the currently most widely
accepted theory for the explosion of a massive star, the
explosion energy is provided by the neutrinos that are abundantly
emitted from the nascent neutron star and interact at a 
probability between 1\% and 10\% with the material of the 
progenitor star. This energy deposition is not only supposed to
power the propagation of the supernova shock into the stellar
mantle and envelope regions and to cause the violent disruption
of the star, but also drives a mass outflow from the surface
of the protoneutron star that continues for more than 10 seconds
and which might be a suitable site for r-process nucleosynthesis
(Woosley \& Hoffman 1992, Witti et al.~1994, Takahashi et al.~1994,
Woosley et al.~1994). The emission of electron lepton 
number and energy via neutrinos determines the evolution of the
hot, collapsed stellar core towards the cold deleptonized neutron
star remnant. Moreover, the interactions of neutrinos with target 
nuclei and nucleons in the neutrino-driven wind and in the stellar 
mantle can have important implications for supernova nucleosynthesis.
Last but not least, the $\sim 20$ neutrinos detected in the 
Baksan (Alexeyev 1988), Kamiokande (Hirata et al.~1987),
and IMB laboratories (Bionta et al.~1987)
in connection with SN~1987A were the
first experimental confirmation of our theoretical picture of 
the events that precede the explosion of a massive star.
Still they serve as a tool to constrain theories of neutrino
properties and particle physics at the extreme conditions of
nascent neutron stars.

As discussed in detail in Lecture~1, the neutrino emission
from the newly born neutron star is characterized by the 
following properties. Electron neutrinos are emitted from the
neutrinosphere with a typical energy of about 10~MeV, while
electron antineutrinos have about 50\% higher energies and
muon and tau neutrinos are radiated with even higher mean
energies. This can be easily be understood by the fact that
$\nu_e$ and $\bar\nu_e$ experience charged-current interactions
while $\nu_{\mu}$ and $\nu_{\tau}$ do not. In addition, due to the 
neutron-proton asymmetry of the medium the coupling
of $\bar\nu_e$ to the medium via absorptions onto protons 
is less strong than $\nu_e$ absorptions onto the more abundant 
neutrons.
Despite the different temperatures of their emission spectra, 
all kinds of neutrinos have roughly similar 
(say, within a factor of 2) luminosities during
most of the Kelvin-Helmholtz cooling phase of the protoneutron 
star. Because the transport of $\nu_{\mu}$ and $\nu_{\tau}$
is strongly affected by isoenergetic scatterings off $n$ and $p$,
their characteristic spectral temperatures and effective 
temperatures are significantly different. The neutrino spectra
are not thermal (``blackbody'') spectra, but are pinched with
a depletion in both the low-energy part and the high-energy
tail. This depletion can be accounted for by using Fermi-Dirac
distributions to describe the emission spectra at a certain time,
and introducing an effective spectral degeneracy parameter
of order unity (typically about 3--5 for $\nu_e$, 2--3 for $\bar\nu_e$,
and between 0 and 2 for $\nu_x \equiv \nu_{\mu},\,\bar\nu_{\mu},\,
\nu_{\tau},\,\bar\nu_{\tau}$).

About 99\% of the total gravitational binding energy that is set free 
during the collapse of the stellar iron core and the formation
of the neutron star is emitted in neutrinos, only about 1\% ends up 
in the kinetic energy of the supernova explosion, and even less,
only about 0.01\% accounts for the spectacularly bright outburst of light
that is seen as type-II supernova event on the sky. Despite the 
enormous amount of energy that is available from the gravitational
collapse, it is not easy to channel the mentioned $\sim 1\%$ or
about $10^{51}\,{\rm erg}$ into kinetic energy. It is still an
unresolved question how type-II manage to do this. It is generally
accepted nowadays that for ``reasonable'' nuclear equations of
state and core masses of the progenitor star the prompt or 
hydrodynamical mechanism does not work: The supernova shock formed
at the moment of core bounce is too weak to overcome the huge 
energy losses due to the disintegration of Fe group nuclei and
additional neutrino losses. The shock stagnates before it reaches
the surface of the stellar Fe core. During the following several
hundred milliseconds of evolution, neutrinos deposit energy behind
the shock. If this neutrino heating is strong enough, the shock
can gain so much energy that it is ``revived'' and can indeed
disrupt the star (see Lecture~1 and the references given there).

However, the neutrino-matter coupling is so weak that this ``delayed''
or neutrino-driven mechanism seems to be extremely sensitive to small 
changes of the physics inside the collapsed stellar core.
In particular, if 
the neutrino fluxes do not surpass a certain threshold value, the 
explosion becomes too weak to be compatible with observations or 
even fizzles. Where are the uncertainties of our current knowledge
of the physics inside stellar iron cores and what might be the missing
link to a stable and robust explosion mechanism? What could help
the neutrino-driven mechanism? Do current models underestimate the
neutrino flux from the neutron star or do they treat the
neutrino-matter coupling in the heating region incorrectly?
\bigskip\medskip\noindent
{\bf 2.$\ $Different possibilities}
\bigskip\smallskip\noindent
There are different lines of exploration currently followed up
by the supernova group at the MPI f\"ur Astrophysik in Garching 
and by other groups 
in Livermore, Los Alamos, and Oak Ridge. On the one hand, these 
activities focus on a closer investigation of the neutrino
interactions and neutrino transport in the protoneutron star.
On the other hand, they concentrate on the simulations of the
hydrodynamical evolution of the collapsed star in more than one
spatial dimension.
\bigskip\noindent
2.1$\ $\underbar{Lower neutrino opacities in the protoneutron star?}
\medskip\noindent
So far, the theoretical understanding of 
neutrino interactions with target nuclei and nucleons
in a dense environment is incomplete and detailed calculations of the
neutrino opacity of a nuclear medium including particle correlation
and screening effects are not yet available. For that reason,
partly also motivated and justified by the effort to simplify the 
numerical
description, all current supernova codes employ neutrino reaction
rates calculated for interactions with isolated target particles.
At least, the rates are more or less reliably 
corrected for blocking effects in the fermion phase spaces. The
real situation may be largely different and more complicated. 

For example, Raffelt \& Seckel (1991) considered auto-correlation
effects for the nucleon spins which can lead to a dramatic reduction
of the axial-vector neutral current (and possibly also charged
current) cross sections. Rapid nucleon spin fluctuations of the
scattering nucleons lead to a reduced effective spin ``seen'' by the
neutrino as a reaction partner. In a parametric study,
Keil et al.~(1995) and 
Janka et al.~(1996) have investigated the effects on the neutrino
cooling of newly formed neutron stars and found a nearly linear
decrease of the cooling time of the protoneutron star with the
global reduction factor for the opacities. This is a non-trivial
result because with reduced opacities also the deleptonisation of
the star is accelerated. As a result of this, the star heats up
faster and to higher peak temperatures. Since the opacities
increase with the temperature (neutrino energy), the net effect
on the cooling of the star is not obvious. Keil et al.~(1995) 
also found that the emitted neutrinos become more energetic when 
the opacity of the protoneutron star is lower and the neutrinosphere
moves deeper into the star. 

The combined effects, reduced cooling
time and increased mean spectral neutrino energies, lead to a 
distortion of the predicted neutrino signal which can be 
compared with the neutrino burst observed in the Kamiokande~II and 
IMB detectors in connection with SN~1987A. Keil et al.~(1995) concluded
that a reduction of the opacities by more than a factor of 2
seems quite unlikely unless the late and low-energy neutrino
events in Kamiokande~2 and IMB are discarded as background
or unless they can be explained by some non-standard neutrino 
emission process, e.g., associated with accretion 
or a spontaneous, first-order phase transition 
in the supranuclear matter that might occur after several seconds of
Kelvin-Helmholtz cooling of the newly born neutron star. 
Despite a lot of vague speculations, there is no
qualified and theoretically founded model for such events in or
at the protoneutron star that can explain the involved energies,
timescales, and structure of the neutrino signal. For a summary 
and discussion of some of the addressed aspects, see Janka (1995).

The derived lower bound for the reduction of the neutrino opacity
which is still compatible with the SN~1987A neutrino signals
sets interesting limits for the nucleon spin fluctuation rate
in the supernova core (Janka et al.~1996).
The theoretical background and formal justification for such
a limit was discussed by Sigl~(1996). Although an opacity
reduction by a factor of about 2 is by far not as much as 
allowed from principle physical reasons, a corresponding doubling
of the neutrino luminosities from the nascent neutron star
would have a very large effect on the supernova explosion.
In order to come to definite conclusions, however, more
reliable calculations of the neutrino opacities in the dense 
medium of supernova cores are urgently called for.
\bigskip\noindent
2.2$\ $\underbar{Convection in the protoneutron star?}
\medskip\noindent
Alternatively, or in addition, convective processes in the 
hot and lepton-rich protoneutron star might raise the neutrino
fluxes and could thus lead to more favorable conditions for   
neutrino heating outside the neutrinosphere and could help
shock revival.

Epstein (1979) pointed out that not only entropy, $S$, inversions
but also zones in the post-collapse core where the lepton fraction,
$Y_l$, decreases with
increasing radius tend to be unstable against Ledoux convection.
Negative $S$ and/or $Y_l$
gradients in the neutrinospheric region and in the layers between the
nascent neutron star and the weakening prompt shock front were realized in
a variety of post-bounce supernova models by Burrows \& Lattimer (1988),
and after shock stagnation
in computations by Hillebrandt (1987) and more recently
by Bruenn (1993), Bruenn \& Mezzacappa (1994),
and Bruenn et al.~(1995). Despite different equations of states
(EOS), $\nu$ opacities, and $\nu$ transport methods,
the development of negative $Y_l$ and $S$ gradients is common
in these simulations and can also be found in protoneutron star 
cooling models of Burrows \& Lattimer (1986), Keil \& Janka (1995), and
Sumiyoshi et al.~(1995).

Convection above the neutrinosphere but below the
neu\-tri\-no-\-heat\-ed
region can hardly be a direct help for the explosion
(Bethe et al.~1987, Bruenn et al.~1995),
whereas convectively enhanced lepton number and energy
transport inside the neutrinosphere
raise the $\nu$ luminosities and can definitely support
neutrino-energized supernova explosions (Bethe et al.~1987).
In this context, Burrows (1987) and Burrows \& Lattimer (1988)
have discussed
entropy-driven convection in the protoneutron star on the basis
of 1D, general relativistic (GR) simulations of the first
second of the evolution of a hot, $1.4\,M_{\odot}$ protoneutron star.
Their calculations were done with a Henyey-like code using a
mixing-length scheme for convective energy and lepton transport.
Recent 2D models (Herant et al.~1994, Burrows et al.~1995,
Janka \& M\"uller 1996 and references therein)
confirmed the possibility that convective
processes can occur in the surface region of the protoneutron star 
immediately
after shock stagnation (``prompt convection'') for a period of at least
several 10~ms. These models, however, have been evolved only over
rather short times or with insufficient numerical resolution in the
protoneutron star or with a spherically symmetrical
description of the core of the protoneutron star that was in some
cases even replaced by an inner boundary condition.

Mayle \& Wilson (1988) and Wilson \& Mayle (1988, 1993)
demonstrated that convection in the nascent neutron star
can be a crucial ingredient that leads to successful delayed explosions.
With the high-density EOS and treatment of the $\nu$
transport used by the Livermore group, however, negative
gradients of $Y_l$ tend to be stabilized by positive
$S$ gradients (see, e.g., Wilson \& Mayle 1989). Therefore
they claim doubly diffusive neutron finger convection to
be more important than Ledoux convection. Doubts about the presence
of doubly diffusive instabilities, on the other hand, were recently
raised by Bruenn \& Dineva (1996).
Bruenn \& Mezzacappa (1994) and Bruenn et al.~(1995) also come to
a negative conclusion about the relevance of prompt convection
in the neutrinospheric region. Although
their post-bounce models show unstable $S$ and $Y_l$
stratifications, the mixing-length approach in their 1D
simulations predicts convective activity inside and around the
neutrinosphere to be present only for 10--30~ms after bounce and to
have no significant impact on the $\nu$ fluxes and spectra when
an elaborate multi-group flux-limited diffusion method is used for
the $\nu$ transfer. Such conclusions
seem to be supported by preliminary 2D simulations
with the same input physics (Guidry 1996). These 2D models,
however, still suffer from the use of an inner boundary condition
at a fixed radius of 20--30~km.

From these differing and partly contradictory
results it is evident that the question whether, where,
when, and how long convection occurs below the neutrinosphere
seems to be a matter of the EOS, of the core structure
of the progenitor star, of the shock properties and propagation,
and of the $\nu$ opacities and the $\nu$ transport description.
In the work which will be reported in Sect.~3.1, 
we compare 1D simulations with the first
self-consistent 2D models that follow the
evolution of the newly formed neutron star for more than a second,
taking into account the GR gravitational
potential and making use of a flux-limited equilibrium diffusion scheme
that describes the transport of $\nu_e$, $\bar\nu_e$, and $\nu_x$
(sum of $\nu_{\mu}$, $\bar\nu_{\mu}$, $\nu_{\tau}$, and $\bar\nu_{\tau}$)
and is very good at high optical depths but only approximative
near the protoneutron star surface (Keil \& Janka 1995).
Our simulations demonstrate that Ledoux convection may continue
in the protoneutron star for a long time and can involve the whole 
star after about one second.
\bigskip\noindent
2.3$\ $\underbar{Convective instabilities in the neutrino-heated region?}
\medskip\noindent
Observations of the light curve and spectra of SN~1987A strongly 
suggest that convective instabilities and aspherical processes
might play an important role not only inside the nascent neutron
star, but also outside of it in its very close vicinity. This is
the region where the radioactive elements, in particular $^{56}$Ni,
which power the supernova light curve, are synthesized
during the explosion. 

The occurrence of large-scale mixing and overturn processes
which must reach deep into the exploding star was clearly indicated by
SN~1987A. From the observations we learned that radioactive material
had been mixed out with very high velocities from the layers
of its formation near the nascent neutron star far into the hydrogen
envelope of the progenitor star. This was supported by the early
detection of X-rays
(Dotani et al.~1987, Sunyaev et al.~1987, Wilson et al.~1988)
and $\gamma$-emission
(Matz et al.~1988, Mahoney et al.~1988, Cook et al.~1988,
Sandie et al.~1988, Gehrels et al.~1988, Teegarden et al.~1989)
and by the observation of strongly Doppler-shifted and -broadened
infrared emission lines of iron group elements
(Erickson et al.~1988, Rank et al.~1988, Barthelmy et al.~1989,
Witteborn et al.~1989, Haas et al.~1990, Spyromilio et al.~1990,
Tueller et al.~1990, Colgan et al.~1994)
at a time
when the photosphere of the supernova was still well inside the hydrogen
envelope. Also, the expanding supernova ejecta developed a
clumpy and inhomogeneous structure quite early during the explosion
(Li et al.~1993).

Besides direct evidence from SN~1987A, theoretical modelling
of the supernova light curve suggested the need for mixing of hydrogen
with a large part of the stellar mantle
(Arnett 1988, Woosley 1988, Shigeyama et al.~1988,
Shigeyama \& Nomoto 1990, Arnett et al.~1989). The smoothness
of the light curve of SN~1987A provided indirect information about
the existence and strength of the mixing process.
Mixing of hydrogen towards the center helps to explain the
smooth and broad light curve maximum by the time-spread of
the liberation of recombination energy. Mixing of heavy
elements into the hydrogen-rich envelope homogenizes the
opacity and again smooths the light curve. Neither the required
amount of mixing nor the observed high velocities of radioactive decay
products could be accounted for merely by Rayleigh-Taylor instabilities
at composition interfaces in the mantle and envelope of the progenitor
star after shock passage
(Arnett et al.~1989; Den et al.~1990; Yamada et al.~1990;
Hachisu et al.~1990, 1991; Fryxell et al.~1991;
Herant \& Benz 1991, 1992).
Moreover, fast moving,
dense explosion fragments outside of the supernova shock front
have recently been discovered in the Vela supernova remnant by
ROSAT X-ray observations
(Aschenbach et al.~1995),
revealing a very clumpy and inhomogeneous
structure of the Vela and other supernova remnants.

In addition, the increasing number of identified high-velocity pulsars
(Harrison et al.~1993, Taylor et al.~1993, Lyne \& Lorimer 1994,
Frail \& Kulkarni 1991, Stewart et al.~1993, Caraveo 1993)
might also be interpreted as an aspect of the new picture
that type-II supernova explosions are by no means spherically symmetrical
events, but that violent processes with noticeable deviation from
spherical symmetry take place in the deep interior of the star during the
early moments of the explosion.

All this was taken as a serious motivation
to extend the modelling of the onset of the explosion to more than one
spatial dimension
(Herant et al.~1992, 1994; Burrows \& Fryxell 1992, 1993; Janka 1993;
Janka \& M\"uller 1993, 1994, 1995a, 1996; M\"uller 1993; 
Miller et al.~1993; Shimizu et al.~1993, 1994; Yamada et al.~1993;
M\"uller \& Janka 1994; Burrows et al.~1995).
These numerical models could indeed show that convective overturn in the
neutrino-heated region around the protoneutron star can be a crucial
help for the explosion. However, it turned out (Janka \& M\"uller 1995a,
Guidry 1996) that even with the helpful effects of convective energy
transport from the heating region towards the supernova shock, the
neutrino energy deposition and thus the neutrino luminosities from
the neutron star have to be larger than some lower threshold. If the
heating is not strong enough, hot-bubble convection does not have
time to develop on the timescale of the accretion of matter from the
shock onto the protoneutron star. In Sect.~3.2 we shall review some
of the results of our simulations.
\bigskip\medskip\noindent
{\bf 3.$\ $Multi-dimensional simulations of convective processes in 
type-II supernovae}
\bigskip\smallskip\noindent
Convection can be driven by a radial gradient of the entropy
per nucleon $S$ and/or by a gradient of the lepton number per
baryon $Y_l$ (Epstein 1979) where $Y_l$ includes contributions
from $e^-$ and $e^+$ and from $\nu_e$ and $\bar\nu_e$ if
the latter are in equilibrium with the matter. Convective
instability in the Ledoux approximation sets in when
$$
{\cal C}_{\rm L}(r)\,\equiv\,
\rund{{\partial \rho\over\partial S}}_{\!\! P,Y_l}
{{\rm d}S\over{\rm d}r} +
\rund{{\partial \rho\over\partial Y_l}}_{\!\! P,S}
{{\rm d}Y_l\over{\rm d}r} \,>\, 0 \ .
\eqno(1)
$$
There are different regions in the collapsed stellar core where
this criterion is fulfilled during different phases of the 
evolution.
\bigskip\smallskip\noindent
3.1$\ $\underbar{Two-dimensional simulations of protoneutron star cooling}
\bigskip\noindent
Simulations of protoneutron star cooling in spherical symmetry
were performed by Burrows \& Lattimer (1986), Suzuki (1989), and
more recently by Keil \& Janka (1995) and 
Sumiyoshi et al.~(1995). These models show the development
of negative gradients of entropy and lepton fraction as the cooling
and deleptonization of the nascent neutron star advances.
Keil et al.~(1996) therefore attempted to do the first
two-dimensional cooling simulations for a period of more than
one second after core bounce.

\midinsert
\hsize = 0.9\hsize
\par\noindent
\epsfxsize=0.75\hsize
\hskip 0.10\hsize
\epsffile{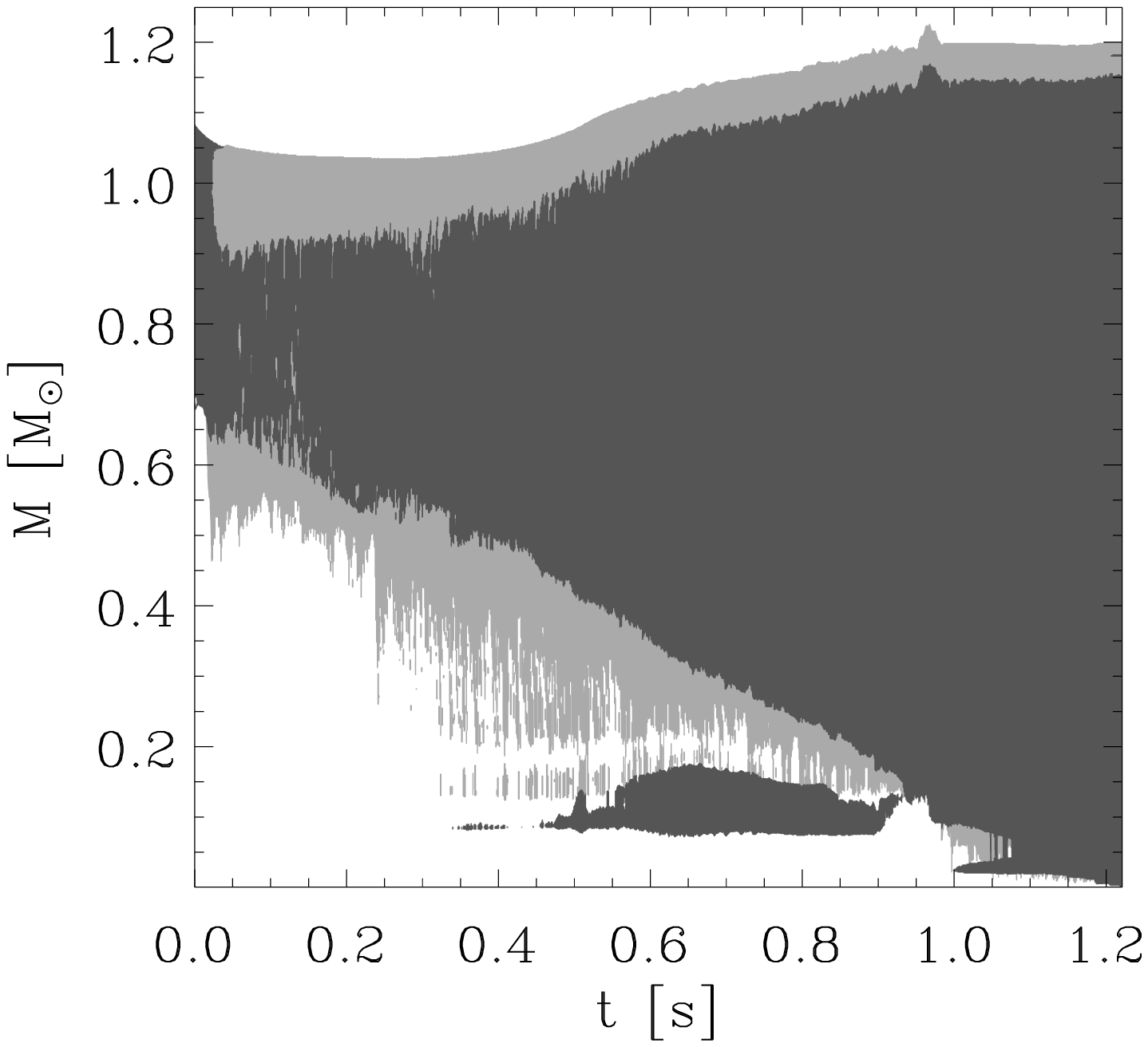}
\vskip 0.25 truecm
\par\noindent
\hskip 0.05\hsize
\vbox{
\hsize = 1.00\hsize
\baselineskip 11.0 pt
\noindent
{\bf Fig.~1.}
Convective (baryon) mass region inside the protoneutron star versus 
time for the 2D simulation. Black indicates regions which are Ledoux
unstable or only marginally stable, grey denotes over- and
undershooting regions where the absolute value of the angular velocity
is $|v_{\theta}| > 10^7\,{\rm cm/s}$.
\vskip 1.00truecm
}
\par\noindent
\par\noindent
\epsfxsize=0.9\hsize
\hskip 0.10\hsize
\epsffile{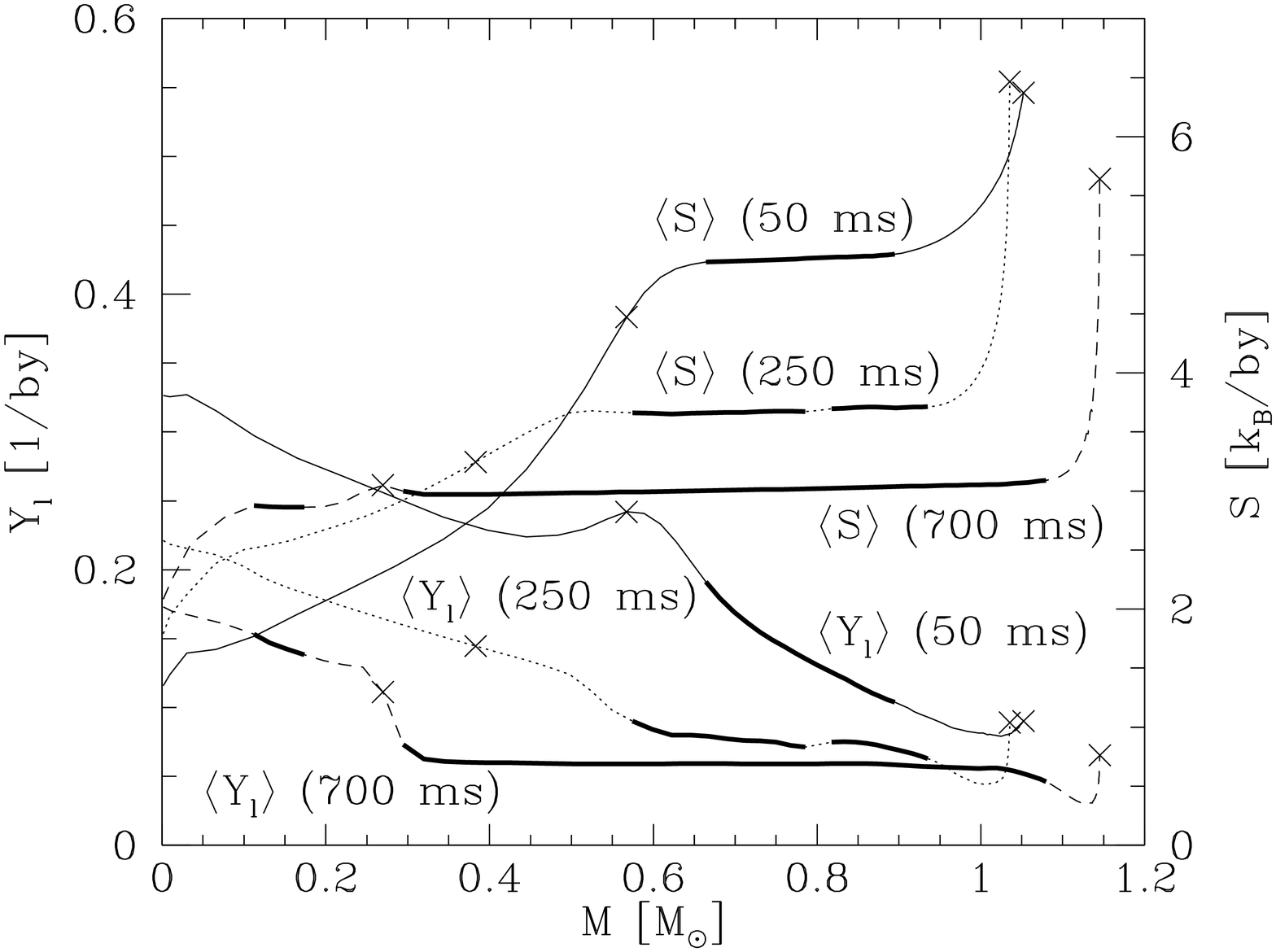}
\vskip 0.25 truecm
\par\noindent
\hskip 0.05\hsize
\vbox{
\hsize = 1.00\hsize
\baselineskip 11.0 pt
\noindent
{\bf Fig.~2.}
Angle-averaged $S$ and $Y_l$ profiles in the protoneutron star.
Thick solid lines indicate regions
that are unstable or only marginally stable against Ledoux convection,
crosses mark boundaries of over- and
undershooting regions where the absolute value of the angular velocity
is $|v_{\theta}| > 10^7\,{\rm cm/s}$.
}
\par\noindent
\endinsert

\bigskip\noindent
{\sl 3.1.1$\ $Numerical implementation}
\medskip\noindent
The simulations were performed with the explicit Eulerian
hydrodynamics code {\it Prometheus} (Fryxell et al.~1989)
that employs a Riemann-solver and is based on the Piecewise
Parabolic Method (PPM) of Colella \& Woodward (1984). A moving grid
with 100 nonequidistant radial zones (initial outer radius
$\sim 60\,{\rm km}$, final radius $\sim 20\,{\rm km}$)
and with up to 60 angular zones was used, corresponding
to a radial resolution of a few 100~m
($\la 1\,{\rm km}$ near the center) and a maximum
angular resolution of $1.5^{\circ}$.
In the angular direction, periodic boundary
conditions were imposed at $\pm 45^{\circ}$ above and below the
equatorial plane. The stellar surface was treated as an
open boundary where the velocity was calculated from the velocity
in the outermost grid zone, the density profile was extrapolated
according to a time-variable power law, and the corresponding
pressure was determined from the condition of hydrostatic
equilibrium. The {\it Prometheus} code was extended for the use of
different time steps and angular resolutions in different
regions of the star. Due to the extremely restrictive
Courant-Friedrichs-Lewy (CFL) condition
for the hydrodynamics, the implicit $\nu$ transport was computed
with typically 10 times larger time steps than the smallest
hydrodynamics time step on the grid
($\sim 10^{-7}\,{\rm s}$) (Keil 1996).

Our simulations were started with the $\sim 1.1\,M_{\odot}$
(baryonic mass) central, dense part
($\rho \ga 10^{11}\,{\rm g/cm}^3$) of the collapsed core of a
$15\,M_{\odot}$ progenitor star (Woosley et al.~1988) that was
computed to a time of about 25~ms after core bounce (i.e., a
few ms after the stagnation of the prompt shock) by Bruenn (1993).
Accretion was not considered but additional matter could be
advected onto the grid through the open outer boundary.
In the 2D run, Newtonian asphericity corrections
were added to the spherically symmetrical GR
gravitational potential: $\Phi_{\rm 2D} \equiv \Phi_{\rm 1D}^{\rm GR}
+ \rund{\Phi_{\rm 2D}^{\rm N}-\Phi_{\rm 1D}^{\rm N}}$. This should
be a sufficiently good approximation because
convective motions produce only local and minor deviations of the
mass distribution from spherical symmetry. Using the GR
potential ensured that transients due to the mapping of
Bruenn's (1993) relativistic 1D results to our code were very small.
When starting our 2D simulation, the radial velocity
(under conservation of the local specific total energy)
was randomly perturbed in the whole protoneutron star with an
amplitude of 0.1\%. The thermodynamics of the neutron star
medium was described by the EOS of Lattimer \& Swesty (1991) which
yields a physically reasonable description of nuclear matter below
about twice nuclear density and is thus suitable to describe the
interior of the considered low-mass neutron star 
($M_{\rm ns}\la 1.2\,M_{\odot}$).

The $\nu$ transport was carried out in radial direction for
every angular zone of the finest angular grid. Angular transport
of neutrinos was neglected. This underestimates the
ability of moving buoyant fluid elements to exchange lepton number
and energy with their surroundings and is only correct if
radial radiative and convective transport are faster. Moreover,
$\nu$ shear viscosity was disregarded. Analytical estimates
(see Keil et al.~1996) show that for typically chosen
numerical resolution the neutrino viscosity is smaller than the
numerical viscosity of the PPM code (which is very small compared
to other hydrodynamics codes), but even the numerical viscosity
is not large enough to damp out the growth and development of
the convective instabilities in the protoneutron star.

\midinsert
\par\noindent
\vbox{
\hsize = 1.00\hsize\noindent
\hskip -1.1truecm
\epsfxsize=10.25truecm \epsffile{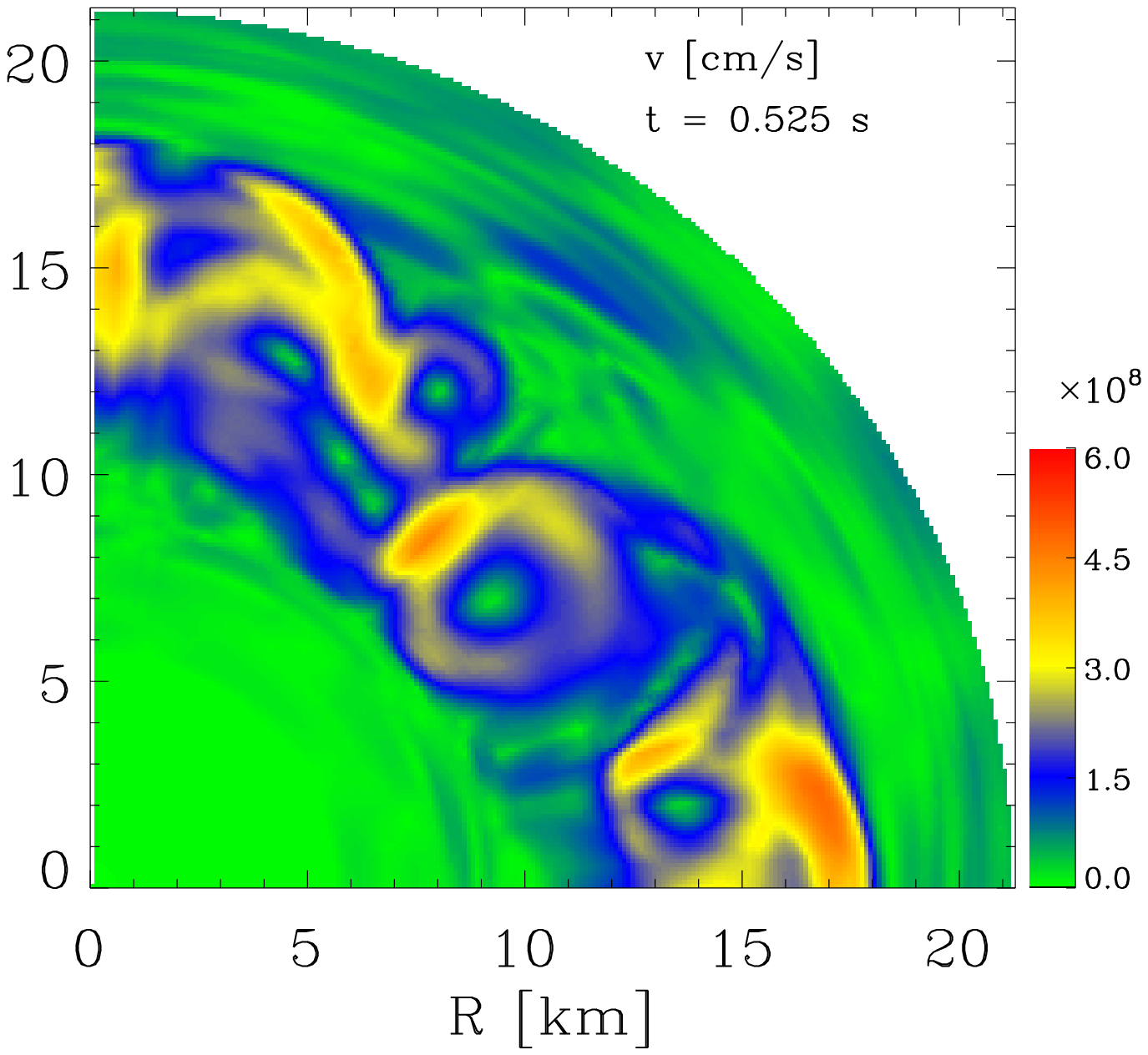}
\hskip -1.3truecm
\epsfxsize=10.25truecm \epsffile{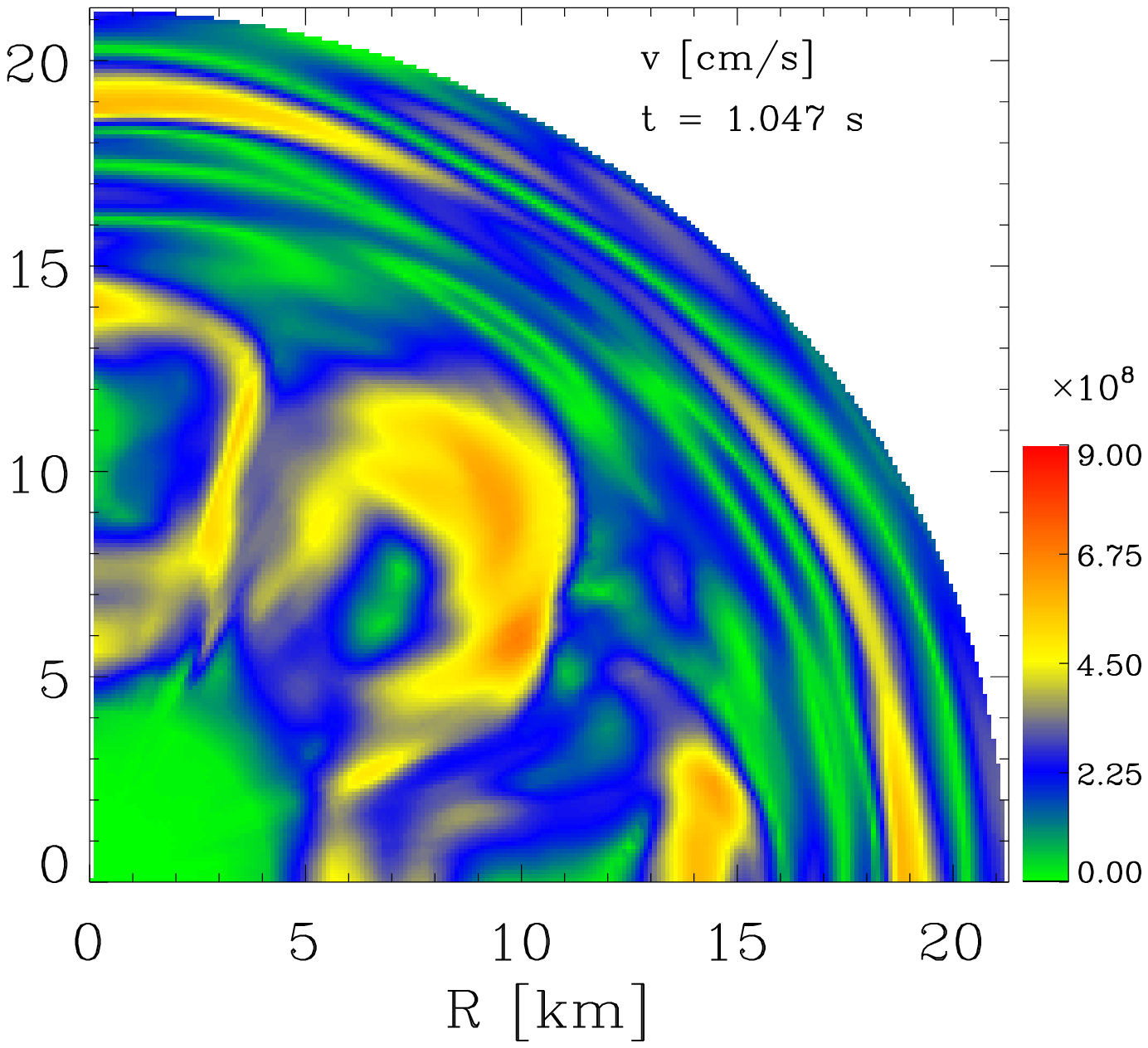}
\noindent
\vskip -8.0truecm
\noindent
\hskip  7.0truecm
{\top a}
\hskip  8.5truecm
{\top b}
\vskip  7.5truecm
\vskip -0.5truecm
\noindent
\hskip -1.1truecm
\epsfxsize=10.25truecm \epsffile{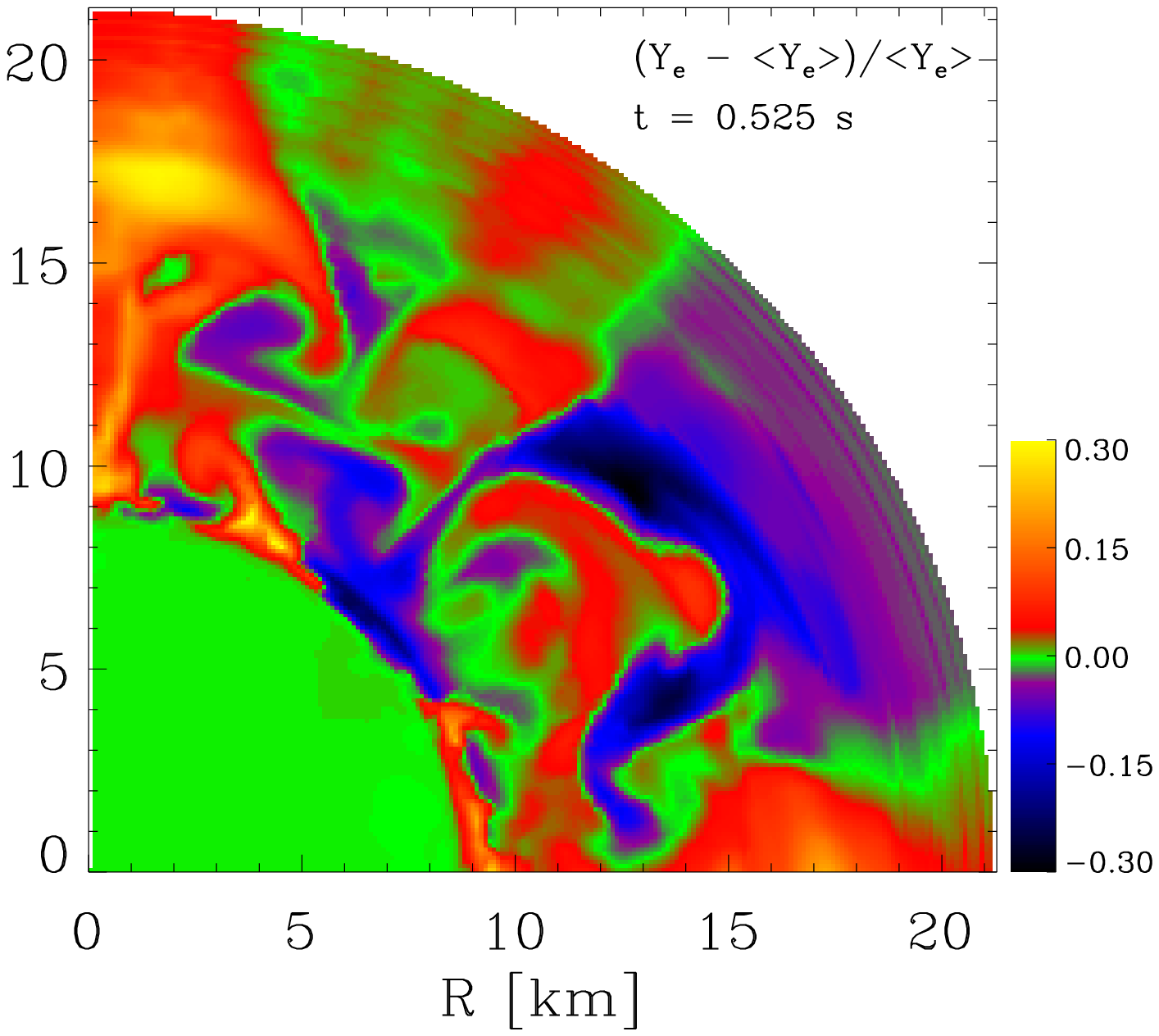}
\hskip -1.3truecm
\epsfxsize=10.25truecm \epsffile{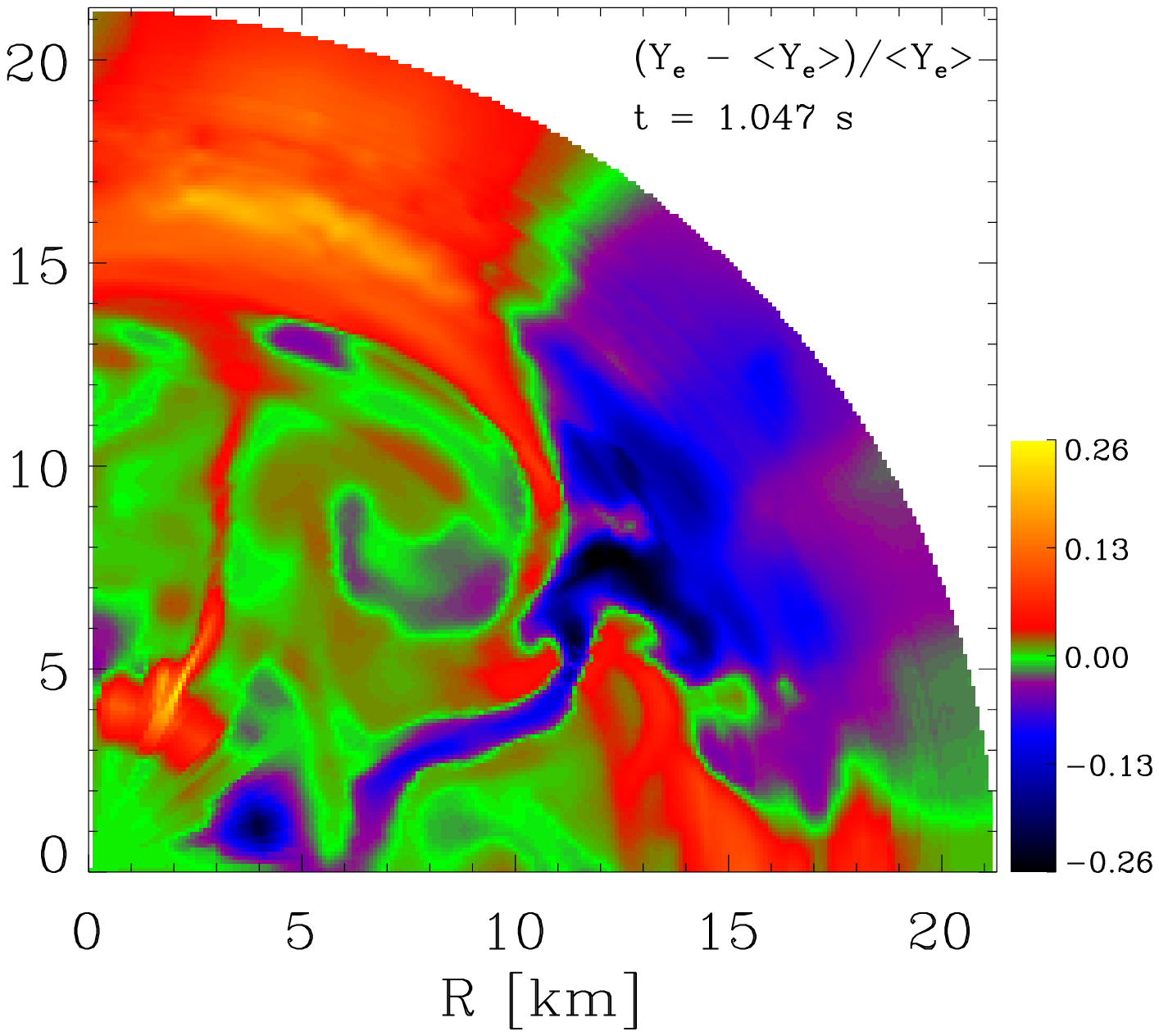}
\noindent
\vskip -8.0truecm
\noindent
\hskip  7.0truecm
{\top c}
\hskip  8.5truecm
{\top d}
\vskip  7.5truecm
}
%
%
%
\hsize = 0.9\hsize
\vskip 0.25 truecm
\par\noindent
\hskip 0.05\hsize
\vbox{
\hsize = 1.00\hsize
\baselineskip 11.0 pt
\noindent
{\bf Fig.~3.}
Panels {\bf a} and {\bf b} show the
absolute values of the velocity for the 2D simulation at times
$t = 0.525~{\rm s}$ and $t = 1.047~{\rm s}$, respectively,
with the grey scale in units of $10^8\,{\rm cm/s}$.
Time is measured from the start of the simulation which is at about
$25\,{\rm ms}$ after core bounce.
The computation was performed in an angular wedge of $90^{\circ}$
between $+45^{\circ}$ and $-45^{\circ}$ around the equatorial
plane. The protoneutron star has contracted to a radius of about
$21\,{\rm km}$ at the given times.
Panels {\bf c} and {\bf d} display the relative deviations of
the electron fraction $Y_e$ from the angular means $\ave{Y_e}$
at each radius for the same two instants. The maximum deviations are
of the order of 30\%. Lepton-rich matter rises while deleptonized
material sinks in. Comparison of both times shows that the
inner edge of the convective layer moves inward from about
$8.5\,{\rm km}$ at $t = 0.525~{\rm s}$ to less than
$2\,{\rm km}$ at $t = 1.047~{\rm s}$.
\vskip 1.0truecm
}
\par\noindent
\endinsert

\bigskip\noindent
{\sl 3.1.2$\ $Results}
\medskip\noindent
Shortly after core bounce, the criterion of Eq.~(1) is fulfilled
between $\sim 0.7\,M_{\odot}$ and $\sim 1.1\,M_{\odot}$ (black area in
Fig.~1) and convective
activity develops within $\sim 10\,{\rm ms}$ after the start of
the 2D simulation. About 30~ms later the outer layers become
convectively stable which is in agreement with
Bruenn \& Mezzacappa (1994). In our 2D simulation, however, the
convectively unstable region
retreats to mass shells $\la 0.9\,M_{\odot}$ and its inner edge
moves deeper into the neutrino-opaque interior of the
star, following a steeply negative lepton
gradient that is advanced towards the stellar center by the
convectively enhanced deleptonization of the outer layers
(Figs.~1 and 2). Note that
the black area in Fig.~1 and the thick solid
lines in Fig.~2 mark not only
those regions in the star which are convectively unstable but also
those {\it which are only marginally
stable} according to the Ledoux criterion of Eq.~(1)
for angle-averaged $S$ and $Y_l$,
i.e., regions where ${\cal C}_{\rm L}(r) \ge
a\cdot\max_r(|{\cal C}_{\rm L}(r)|)$
with $a = 0.05$ holds. For $a\la 0.1$ the accepted region varies
only little with $a$ and is always embedded by the grey-shaded
area where the absolute value of the angular velocity is
$|v_{\theta}| > 10^7\,{\rm cm/s}$.
Yet, only sporadically and randomly appearing patches
in the convective layer fulfill Eq.~(1) rigorously.
Figure~2 shows that the black region
in Fig.~1 coincides with the layers
where convective mixing flattens the $S$ and $Y_l$ gradients.

\midinsert
\hsize = 0.9\hsize
\par\noindent
\epsfxsize=0.9\hsize
\hskip 0.10\hsize
\epsffile{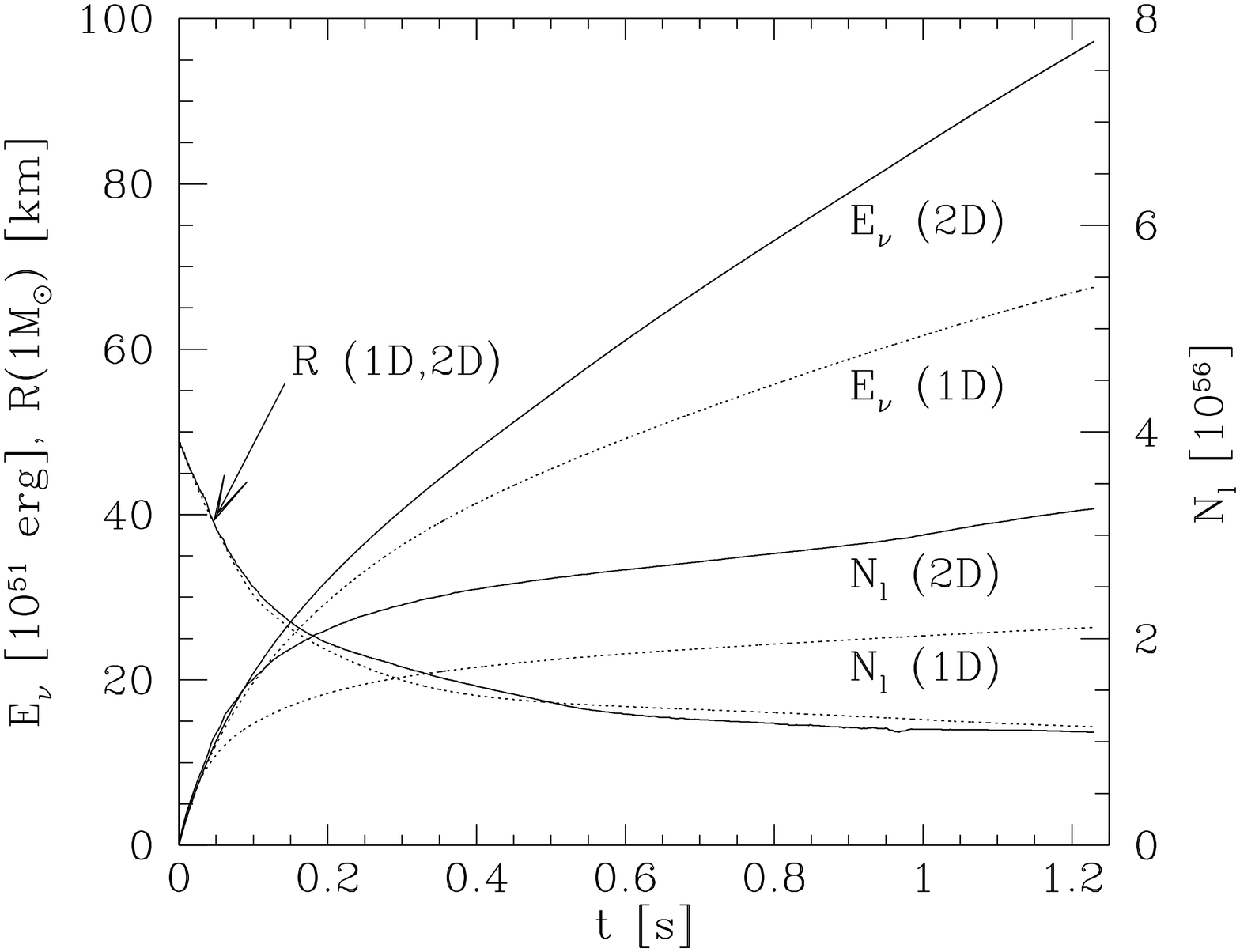}
\vskip 0.25 truecm
\par\noindent
\hskip 0.05\hsize
\vbox{
\hsize = 1.00\hsize
\baselineskip 11.0 pt
\noindent
{\bf Fig.~4.}
Radius of the $M = 1\,M_{\odot}$ mass shell and
total lepton number $N_{\rm l}$ and energy
$E_{\nu}$ radiated away by neutrinos vs.~time for
the 2D (solid) and 1D (dotted) simulations.
\vskip 1.25truecm
}
\par\noindent
\par\noindent
\epsfxsize=0.9\hsize
\hskip 0.10\hsize
\epsffile{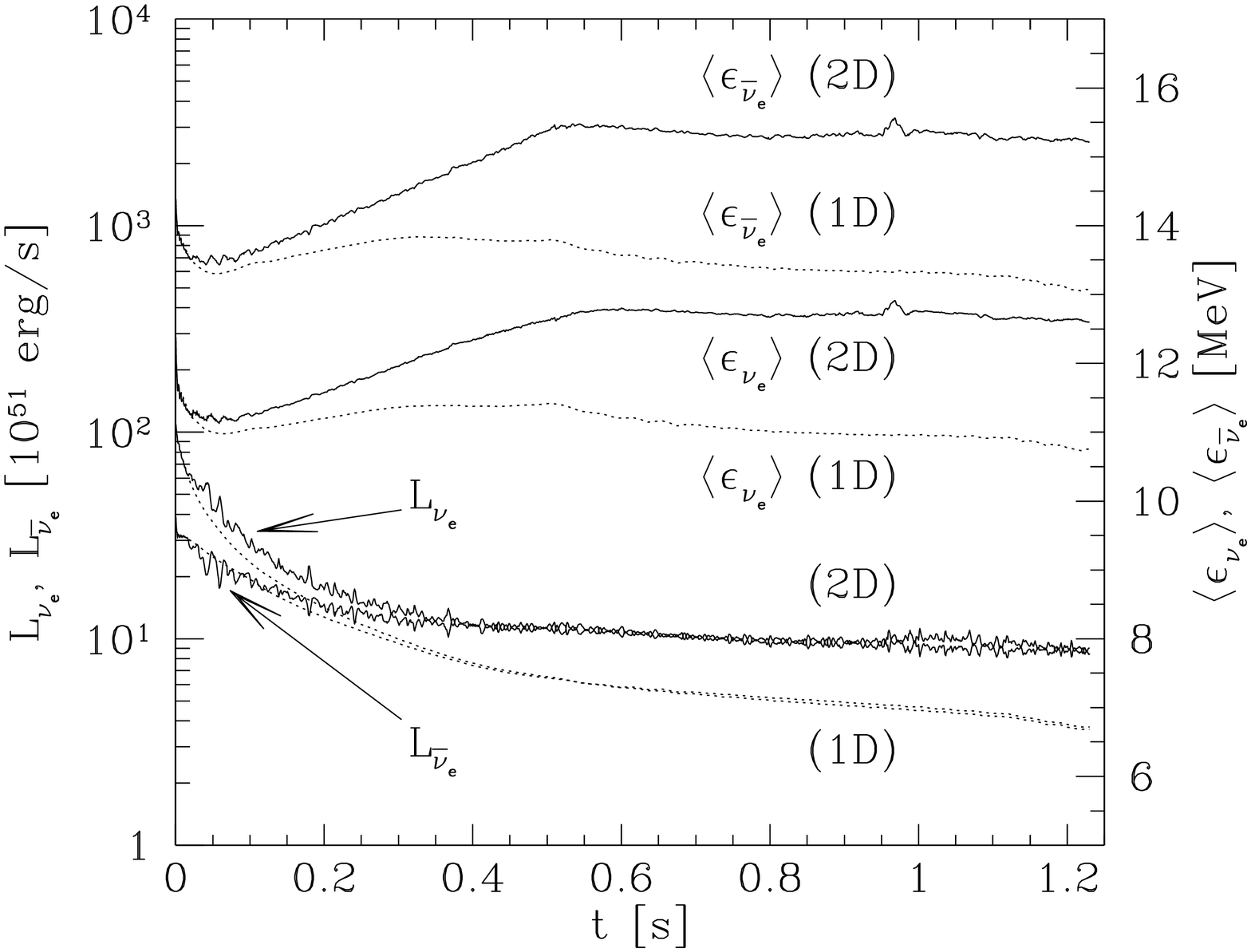}
\vskip 0.25 truecm
\par\noindent
\hskip 0.05\hsize
\vbox{
\hsize = 1.00\hsize
\baselineskip 11.0 pt
\noindent
{\bf Fig.~5.}
$\nu_e$ and $\bar\nu_e$ luminosities and
mean energies vs.~time for the 2D
simulation (solid) compared with the 1D
run (dotted).
\vskip 1.0truecm
}
\par\noindent
\endinsert

The convective pattern is extremely non-stationary and has
most activity on large scales with radial coherence lengths of
several km up to $\sim 10\,{\rm km}$ and convective ``cells''
of 20$^{\circ}$--30$^{\circ}$ angular diameter, at some times
even 45$^{\circ}$ (Fig.~3). Significant over-
and undershooting takes place (grey regions
in Fig.~1) and the convective mass motions
create pressure waves and perturbations in the
convectively stable neutron star interior and in the surface 
layers. The maximum
convective velocities are usually $\sim 4\cdot 10^8\,{\rm cm/s}$, but
peak values of $\sim 10^9\,{\rm cm/s}$ can be reached. These
velocities are typically 5--10\% of the average sound speed
in the star. The kinetic energy of the convection is several
$10^{49}\,{\rm erg}$ at $t\la 1\,{\rm s}$ and climbs to
$\sim 2\cdot 10^{50}\,{\rm erg}$ when the protoneutron star
is fully convective.
Relative deviations of $Y_l$ from the angular mean
can be several 10\% (even 100\%) in rising or sinking buoyant
elements, and for $S$ can reach 5\% or more. Rising flows
always have larger $Y_l$ {\it and} $S$ than their surroundings.
Corresponding
temperature and density fluctuations are only $\sim 1$--3\%.
Due to these properties and the problems in applying the Ledoux
criterion with angle-averaged $S$ and $Y_l$
straightforwardly, we suspect that it is hardly possible
to describe the convective activity with a mixing-length treatment
in a 1D simulation.

Our 2D simulation shows that convection in the
protoneutron star can encompass
the whole star within $\sim 1\,{\rm s}$ and can continue for at least
as long as the deleptonization takes place, possibly even longer.
A deleptonization ``wave'' associated with the convectively
enhanced transport moves towards the center of the protoneutron
star. This reduces the timescale for the electron fraction $Y_e$ to
approach its minimum central value of about 0.1 from
$\sim 10\,{\rm s}$ in the 1D case, where the lepton loss proceeds
much more gradually and coherently, to only $\sim 1.2\,{\rm s}$
in 2D.
With convection the entropy and temperature near the center rise
correspondingly faster despite a similar contraction of the star
in 1D and 2D
(Fig.~4). Convection increases the total lepton
number flux and the $\nu$ luminosities by up to a factor of 2
(Fig.~5) and therefore the emitted lepton number
$N_l$ and energy $E_{\nu}$ rise much more rapidly
(Fig.~4). The convective energy (enthalpy
plus kinetic energy) flux dominates the diffusive $\nu$ energy flux
in the convective mantle after $t \ga 250\,{\rm ms}$ and becomes
more than twice as large later. Since convection takes place somewhat
below the surface, $\nu$'s take over the energy transport
exterior to $\sim 0.9\,M_{\odot}$. Thus the surface $\nu$ flux shows
relative anisotropies of only 3--4\%, in peaks up to $\sim 10\%$,
on angular scales of 10$^{\circ}$--40$^{\circ}$.
Averaged over all directions,
the neutrinospheric temperatures and mean energies
$\ave{\epsilon_{\nu_i}}$ of the emitted $\nu_e$ and $\bar\nu_e$
are higher by 10--20\% (Fig.~5).

\bigskip\noindent
{\sl 3.1.3$\ $Consequences}
\medskip\noindent
Convectively increased neutrino emission from the protoneutron
star does not only have influence on the supernova explosion
mechanism. It is also of crucial importance to understand the 
nucleosynthesis in the neutrino-heated supernova ejecta 
and in the neutrino-driven wind whose
degree of neutronization is determined by the interaction with
the $\nu_e$ and $\bar\nu_e$ fluxes. Absorptions of $\nu_e$ onto
free neutrons increase the free proton abundance while captures of 
$\bar\nu_e$ onto protons make the matter more $n$-rich.

As discussed above, convective neutrino transport in the 
nascent neutron star accelerates the deleptonization of the
protoneutron star drastically. This means that during the first
second or so the $\nu_e$ number flux ${\cal N}_{\nu_e}$
is increased relative to the $\bar\nu_e$ number flux
${\cal N}_{\bar\nu_e}$. If the protoneutron star atmosphere 
(where the neutrinospheres are located) is not convective but
a radiative layer sitting on top of a convective region,
then it can be shown (Keil et al.~1996)
that the ratio of the average energies of $\bar\nu_e$ and $\nu_e$,
$\ave{\epsilon_{\bar\nu_e}^n}/\ave{\epsilon_{\nu_e}^n}$ ($n$ is
an arbitrary power), is not influenced very much by the convective
activity deeper inside the star. In that case it is easy to
see that the electron fraction in the ejecta,
$Y_e^{\rm ej}\approx 
1/\eck{1+({\cal N}_{\bar\nu_e}\ave{\epsilon_{\bar\nu_e}^2})/
({\cal N}_{\nu_e}\ave{\epsilon_{\nu_e}^2})}$ (Qian \& Woosley 1996),
will increase when the ratio ${\cal N}_{\bar\nu_e}/{\cal N}_{\nu_e}$
decreases. Keil et al.~(1996) found that this
is indeed the case before about 0.4~s after shock formation.
This effect offers a solution of the overproduction problem
of $N = 50$ nuclei in current supernova models (see
Hoffman et al.~1996, McLaughlin et al.~1996).

Because the opacity of the protoneutron star increases for 
$\nu_e$ (which are absorbed on neutrons) but decreases for
$\bar\nu_e$ (absorbed on protons) with progressing neutronization
of the matter, the accelerated neutronization of the convective
protoneutron star will lead to a more rapid increase of
$\ave{\epsilon_{\bar\nu_e}}$ relative to $\ave{\epsilon_{\nu_e}}$
than in 1D models at times later than about 1~s.
This will favor a faster drop of $Y_e^{\rm ej}$ and thus might
help to produce the $n$-rich conditions required for a possible
r-processing in the high-entropy neutrino-driven wind (for
details, see Woosley et al.~1994, Takahashi et al.~1994).

Without any doubt, an increase of the neutrino luminosities 
by a factor of $\sim 2$ during the first second after core
bounce may be decisive for a successful explosion via the
neutrino-heating mechanism. This will have to be investigated
in future multi-dimensional simulations where not only the 
evolution of the protoneutron star is followed but the 
whole collapsed star is included. Parametric 1D and 2D studies
with varied neutrino luminosities carried out by 
Janka \& M\"uller (1995a, 1996) have already 
demonstrated the sensitivity of the explosion to changes of
the luminosity of the order of some 10\%. In the next section
these results will be addressed.
\bigskip\smallskip\noindent
3.2$\ $\underbar{Two-dimensional simulations of convection in the 
neutrino-heated region}
\bigskip\noindent
Herant et al.~(1992) first demonstrated by a hydrodynamical
simulation that strong, turbulent overturn occurs in the
neutrino-heated layer outside of the protoneutron star and
that this helps the stalled shock front to start
re-expansion as a result of energy deposition by neutrinos.
Although the existence and fast growth of these
instabilities was confirmed by Janka \& M\"uller (1994, 1995a, 1996)
the results of their simulations in 1D and in 2D indicated
a very strong sensitivity to the conditions at the
protoneutron star and to the details of the description
of neutrino interactions and neutrino transport. Since the
knowledge about the high-density equation of state in the
nascent neutron star and about the neutrino opacities
of dense matter is incomplete (see Sect.~2.1),
the influence of a contraction
of the neutron star and of the size of the neutrino fluxes
on the evolution of the explosion has been tested by
systematic studies.

In the following we shortly report the main conclusions that
can be drawn from our set of 1D and 2D models with different
core-neutrino luminosities and with varied temporal
contraction of the inner boundary (Janka \& M\"uller 1996).
\bigskip\noindent
{\sl 3.2.1$\ $Numerical implementation}
\medskip\noindent
The inner boundary was placed
somewhat inside the neutrinosphere and was used instead of
simulating the evolution of the very dense inner core of
the nascent neutron star. This gave us the freedom to set
the neutrino fluxes to chosen values at the inner boundary
and also enabled us to follow the 2D simulations until about
one second after core bounce with a reasonable number
(${\cal O}(10^5)$) of time steps and an ``acceptable'' computation
time, i.e.~several 100$\,$h on one processor of a Cray-YMP with
a grid of $400\times 90$ zones and a highly efficient implementation
of the microphysics. Note that doubling the angular resolution
multiplies the computational load by a factor of about 4!

Our simulations started at $\sim 25\,{\rm ms}$ after
shock formation from an initial model evolved
through core collapse and bounce by Bruenn (1993).
Boundary motion, luminosities of all neutrino kinds, and
non-thermal neutrino spectra were time-dependent and mimiced the behavior
in Bruenn (1993) and in Newtonian computations
by Bruenn et al.~(1995). Except for Doppler-shift and gravitational
redshift the neutrino fluxes were kept constant with radius and did not
include accretion luminosity. Neutrinos interact with matter
by scattering on $e^\pm$, $n$, $p$, $\alpha$, and nuclei,
by neutrino pair processes, and $\nu_e$ and $\bar\nu_e$ also by the
$\beta$-reactions. The reaction rates were evaluated by using
Monte Carlo calibrated variable Eddington factors that depended on the
density gradient at the neutron star surface. Inside the neutrinosphere
reactive equilibrium between neutrinos and matter can be established.
Our EOS described $e^\pm$ as arbitrarily relativistic,
ideal Fermi gases and $n$, $p$, $\alpha$, and a representative
nucleus as ideal Boltzmann gases in NSE
(good at $\rho\la 5\times 10^{13}\,{\rm g/cm}^3$ for temperatures
$T\ga 0.5\,{\rm MeV}$). 2D computations were performed with the
Eulerian {\it Prometheus} code 
(up to 1$^{\rm o}$ resolution and $400\times 180$ zones),
1D runs with a Lagrangian method
(details in Janka \& M\"uller 1996).

\bigskip\noindent
{\sl 3.2.2$\ $Results for spherically symmetric models}
\medskip\noindent
The evolution of the stalled, prompt shock in 1D
models turned out to be extremely sensitive to the size of the
neutrino luminosities and to the corresponding strength of
neutrino heating exterior to the gain radius. In models with
successively higher core neutrino fluxes the shock is driven
further and further out to larger
maximum radii during a phase of $\sim 100$--$150\,{\rm ms}$ of
slow expansion. Nevertheless, it finally recedes again
to become a standing accretion shock at a much smaller radius
(Fig.~6, dotted lines; see also Fig.~8).
For a sufficiently high threshold luminosity, however,
neutrino heating is strong enough to
cause a successful explosion (Fig.~6, solid lines). For even
higher neutrino fluxes the explosion develops faster and gets
more energetic. In case of our $15\,M_\odot$ star with
$1.3\,M_\odot$ Fe-core (Woosley et al.~1988)
we find that explosions occur for $\nu_e$ and $\bar\nu_e$
luminosities above
$2.2\times 10^{52}\,{\rm erg/s}$ in case of a contracting
inner boundary
(to mimic the shrinking protoneutron star), but of only
$1.9\times 10^{52}\,{\rm erg/s}$ when the radius of the inner
boundary is fixed.

\midinsert
\hsize = 0.9\hsize
\par\noindent
\epsfxsize=0.9\hsize
\hskip 0.10\hsize
\epsffile{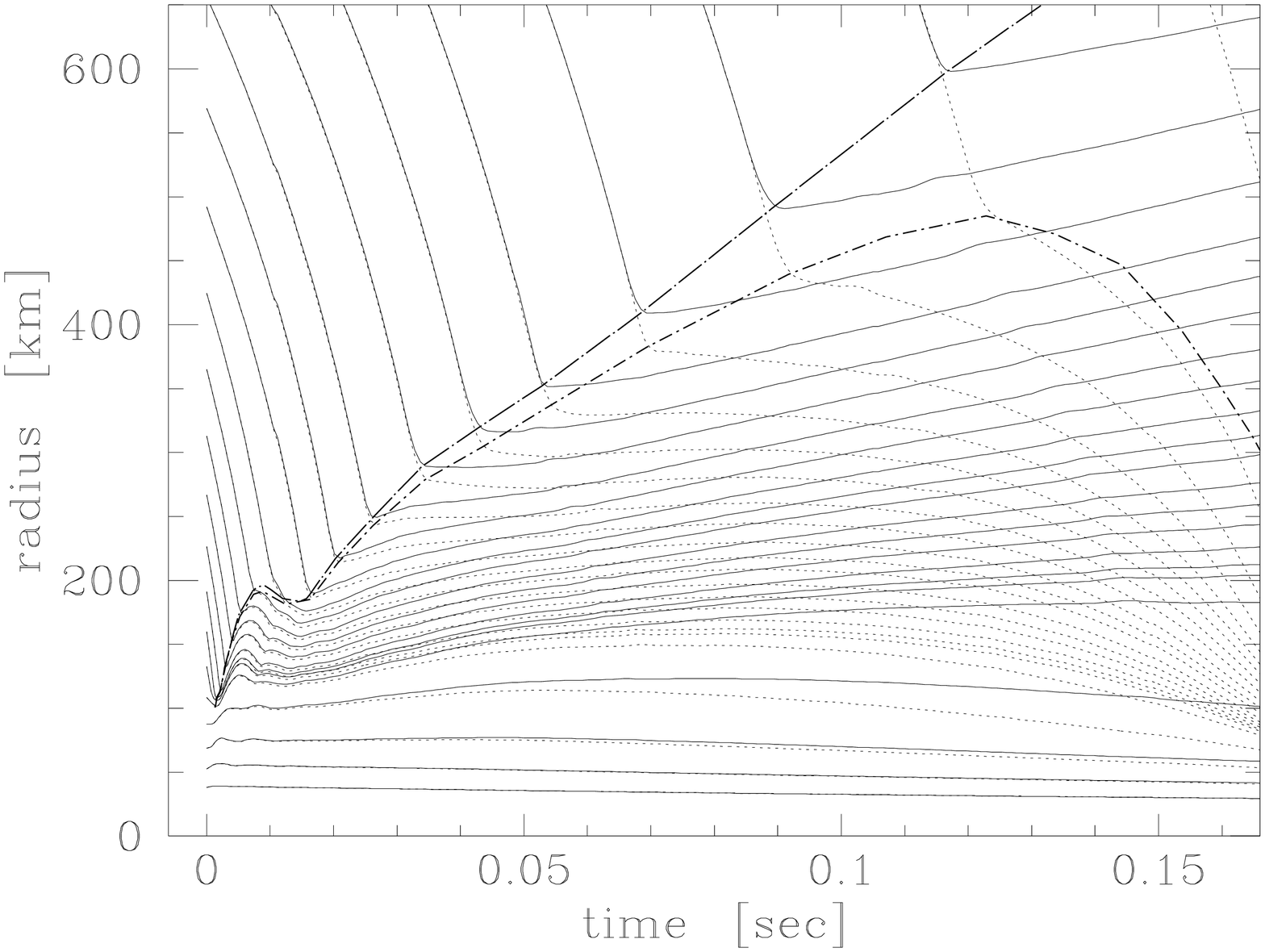}
\vskip 0.25 truecm
\par\noindent
\hskip 0.05\hsize
\vbox{
\hsize = 1.00\hsize
\baselineskip 11.0 pt
\noindent
{\bf Fig.~6.}
Radial positions of mass shells of a marginally exploding
model (solid lines) with initial neutrino luminosities 
from the lower boundary of $2.2\cdot 10^{52}\,{\rm erg/s}$
and of a still unsuccessful 1D model
(dotted lines; initial boundary luminosities 
$2.1\cdot 10^{52}\,{\rm erg/s}$) versus time after core bounce.
The thick broken lines mark the shock positions.
}
\par\noindent
\endinsert

The transition from failure to explosion requires the
neutrino luminosities to exceed some threshold value. Yet,
this is not sufficient. High neutrino energy deposition
has to be maintained for a longer period of time to ensure high
pressure behind the shock. If the decay of the neutrino fluxes is
too fast, e.g., if a significant fraction of the
neutrino luminosity comes from neutrino emission by spherically accreted
matter, being shut off when the shock starts to expand, then the
outward shock propagation may break down again and the
model fizzles. Continuous shock expansion needs a
sufficiently strong push from the neutrino-heated
matter until the material
behind the shock has achieved escape velocity and does not need
pressure support to make its way out.

This contradicts a recent suggestion by Burrows \& Goshy (1993)
that the explosion can be viewed at as a
global instability of the star that, once excited, inevitably leads
to an explosion. The analysis by Burrows \& Goshy may allow one to
estimate the radius of shock stagnation when stationarity applies.
The start-up phase of the explosion, however, can
hardly be described by steady-state assumptions, because
the timescales of shock expansion, of
neutrino cooling and heating, and of temperature and density changes
between neutron star and shock are all of the
same order, although long compared to the sound crossing
time and (possibly) shorter than the characteristic
times of luminosity changes and variations of the mass accretion
rate into the shock.
In particular, due to the high sound speed and rather
slow shock expansion the shock propagation is very sensitive
to changes of the conditions in the neutrino-heated layer.
A contraction of the neutron star or enhanced cooling of the gas
inside the gain radius accelerate the advection of matter
through the gain radius and reduce the time the postshock material
is heated. This is harmful to the outward motion of the shock
just as a moderate decline of the neutrino fluxes can be.

\midinsert
\hsize = 0.9\hsize
\par\noindent
\epsfxsize=0.9\hsize
\hskip 0.10\hsize
\epsffile{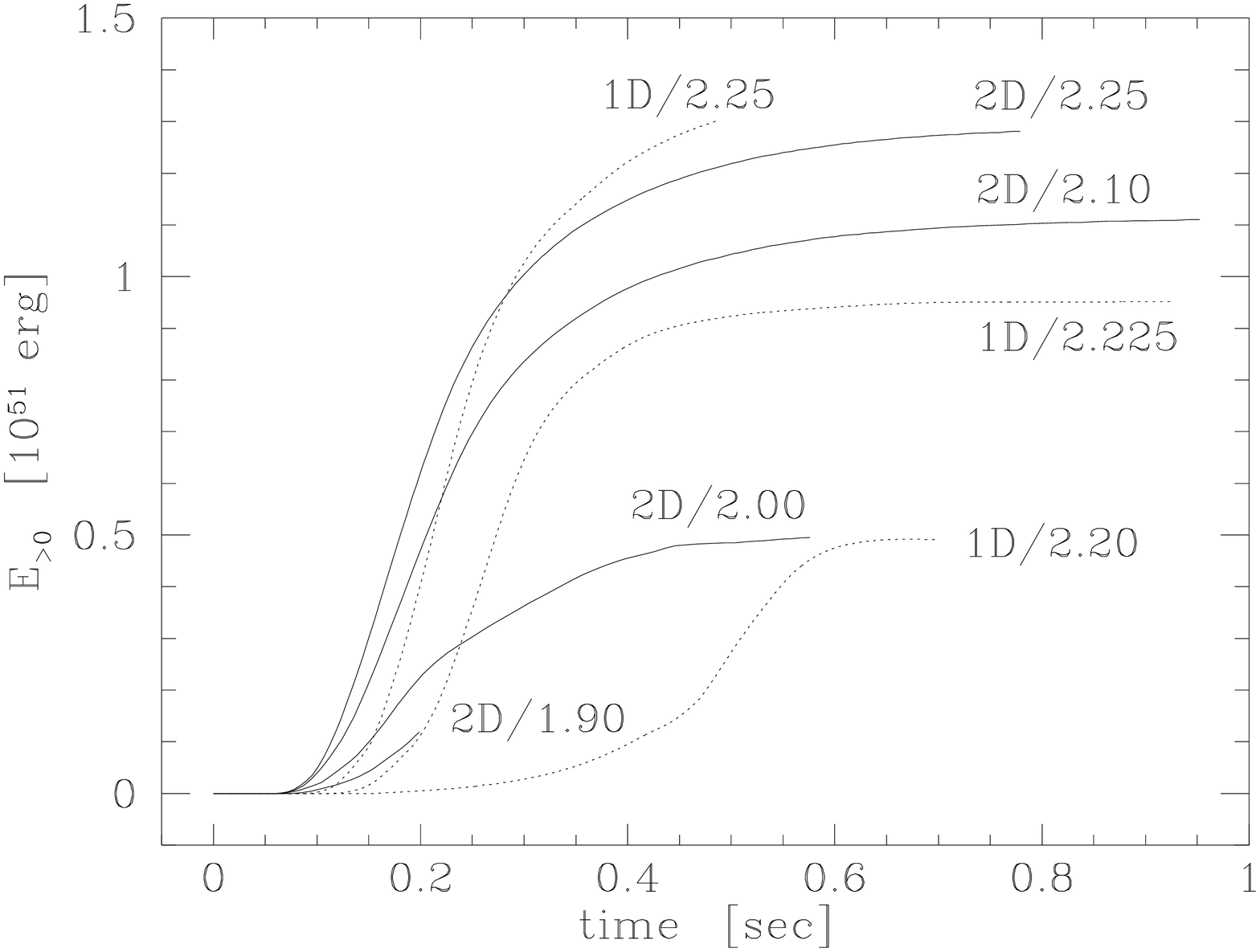}
\vskip 0.25 truecm
\par\noindent
\hskip 0.05\hsize
\vbox{
\hsize = 1.00\hsize
\baselineskip 11.0 pt
\noindent
{\bf Fig.~7.}
Explosion energies vs.~time after the start of
the simulations ($\sim 25\,$ms after bounce)
for exploding 1D (dotted) and 2D models (solid).
The numbers denote the initial $\nu_e$ and
$\bar\nu_e$ (approximately) luminosities in 10$^{52}\,$erg/s.
\vskip 0.60truecm
}
\par\noindent
\par\noindent
\epsfxsize=0.9\hsize
\hskip 0.10\hsize
\epsffile{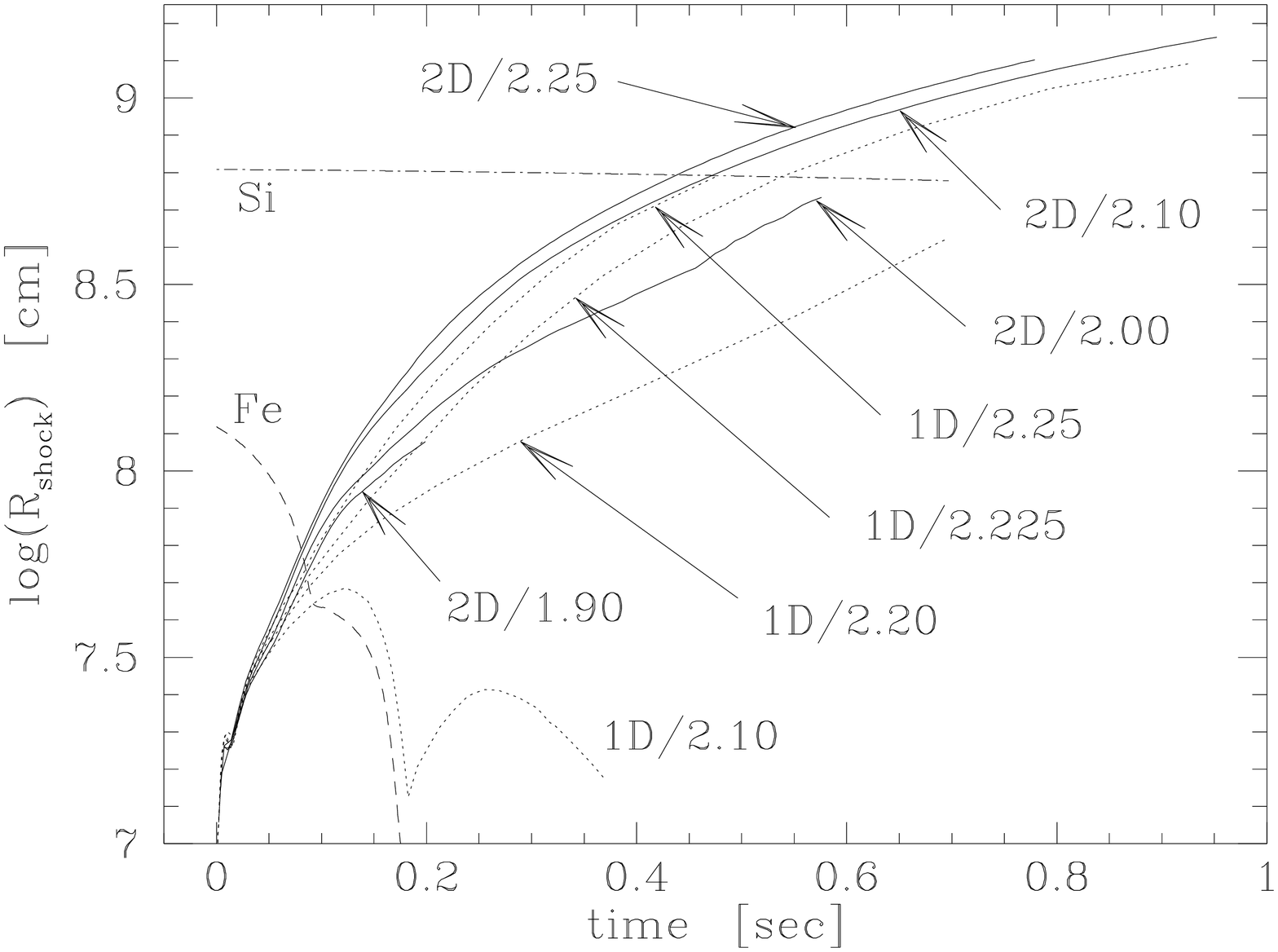}
\vskip 0.25 truecm
\par\noindent
\hskip 0.05\hsize
\vbox{
\hsize = 1.00\hsize
\baselineskip 11.0 pt
\noindent
{\bf Fig.~8.}
Shock positions vs.~time after core bounce
for the exploding one-dimensional (``1D'', dotted lines) and 
two-dimensional (``2D'', solid lines) models of Fig.~7. 
The numbers indicate the size of the initial $\nu_e$ (and
$\bar\nu_e$, approximately) 
luminosities in units of $10^{52}\,{\rm erg}/{\rm s}$.
In addition, the result for
the unsuccessful 1D model with initial boundary luminosities
of $2.1\cdot 10^{52}\,{\rm erg/s}$ is depicted. 
The dashed and dash-dotted curves mark the mass shells
that correspond to the outer boundaries of Fe-core and
Si-shell, respectively, for the latter model.
}
\par\noindent
\endinsert
\bigskip\noindent
{\sl 3.2.3$\ $Results for two-dimensional models}
\medskip\noindent
In spherical symmetry the expansion of the neutrino-heated
matter and of the shock can occur only when also the overlying material
is lifted in the gravitational field of the neutron star.
In the multi-dimensional case this is different. Blobs and
lumps of heated matter can rise by pushing colder material
aside and cold material from the region behind the shock
can get closer to the zone of strongest neutrino heating
to readily absorb energy. Also, when buoyancy forces drive
hot matter outward, the energy loss by re-emission of neutrinos
is significantly reduced. Thus overturn of low-entropy and
high-entropy gas increases the
efficiency of neutrino energy deposition outside the radius of 
net energy gain and leads to explosions in 2D already for lower
neutrino fluxes than in the spherically symmetrical case.
Our models, however, do not show the existence of a ``convective cycle"
or ``convective engine" (Herant et al.~1994)
that transports energy from the heating region into the shock.
The matter between protoneutron star and shock is subject
to strong neutrino heating and cooling and our high-resolution
calculations reveal a turbulent, unordered, and
dynamically changing pattern of
rising and sinking lumps of material with very different
thermodynamical conditions and no clear indication of
inflows of cool
gas and outflows of hot gas at well-defined thermodynamical
states.

2D models explode for core neutrino luminosities
which cannot produce explosions in 1D. There is a window of
neutrino fluxes with a width of $\sim 20$\% of the
threshold luminosity for explosions in 1D, where convective
overturn between gain radius and shock is a
significant help for shock revival. For lower
neutrino fluxes even convective overturn cannot ensure strong
explosions but the explosion energy gets very low. We do not
find a continuous ``accumulation" (Herant et al.~1994)
of energy in the convective
shell until an explosion energy typical of a type-II supernova
is reached. For neutrino fluxes that cause powerful explosions
already in 1D,
turbulent overturn occurs but is not crucial for the explosion.
In fact, in this case
the fast rise of bubbles of
heated material leads to a less vigorous start of the explosion and
to the saturation of the explosion energy at a somewhat lower
level (Fig.~7).
The explosion energy, defined as the {\it net energy of the
expanding matter at infinity}, does not exceed $10^{50}\,{\rm erg}$
earlier than after $\sim 100\,{\rm ms}$ of neutrino heating.
This is the characteristic timescale of neutrinos to transfer
an amount of energy
to the material that is roughly equal to its gravitational
binding energy and it is also the timescale that the
convective overturn between gain radius and shock needs to develop
to its full strength.
It is not possible to determine or predict the final explosion energy
of the star from a short period of only 100--$200\,{\rm ms}$
after shock formation. Typically, the increase of the
explosion energy with time levels off not before
400--$500\,{\rm ms}$
after bounce, followed by only a very slow increase due
to the much smaller contributions of the few $10^{-3}\,M_\odot$
of matter blown away from the
protoneutron star in the neutrino wind (Fig.~7).
Since the wind material is heated slowly and can expand
as soon as the internal energy per nucleon roughly
equals its gravitational binding energy, the matter does not have
a large kinetic energy at infinity.

Although the global evolution of powerful explosions in 2D,
i.e., the increase of the explosion energy with time (Fig.~7),
the shock radius as a function of time (Fig.~8),
or even the amount of $^{56}$Ni produced by explosive
nucleosynthesis, is not much different from energetic
explosions of spherically symmetrical models, the structure of the
shock and of the thick layer of expanding, dense matter behind
the shock clearly bear the effects of the turbulent activity.
The shock is deformed on large scales and its expansion velocity
into different directions varies by $\sim 20$--30\%. The material
behind the shock reveals large-scale inhomogeneities in density,
temperature, entropy, and velocity,
these quantities showing contrasts of up to a factor of 3.
The typical angular scale of the largest structures is
$\sim 30^{\rm o}$--45$^{\rm o}$. We do not find
that the turbulent pattern tries to gain power
on the largest possible scales and to evolve into the
lowest possible mode, $l=1$ (Herant et al.~1992, 1994).
Turbulent motions are still going on in the extended, dense
layer behind the shock
when we stop our calculations at $\sim 1\,{\rm s}$ after
bounce. We consider them as possible origin of the anisotropies,
inhomogeneities, and non-uniform distribution of radioactive
elements which were observed in SN~1987A. The contrasts
behind the shock are about an order of magnitude
larger than the artificial perturbations that were
used in hydrodynamical simulations to trigger the growth of
Rayleigh-Taylor instabilities in the stellar mantle and envelope.

\midinsert
\hsize = 0.9\hsize
\par\noindent
\epsfysize=20.0truecm
\hskip 0.10\hsize
\epsffile{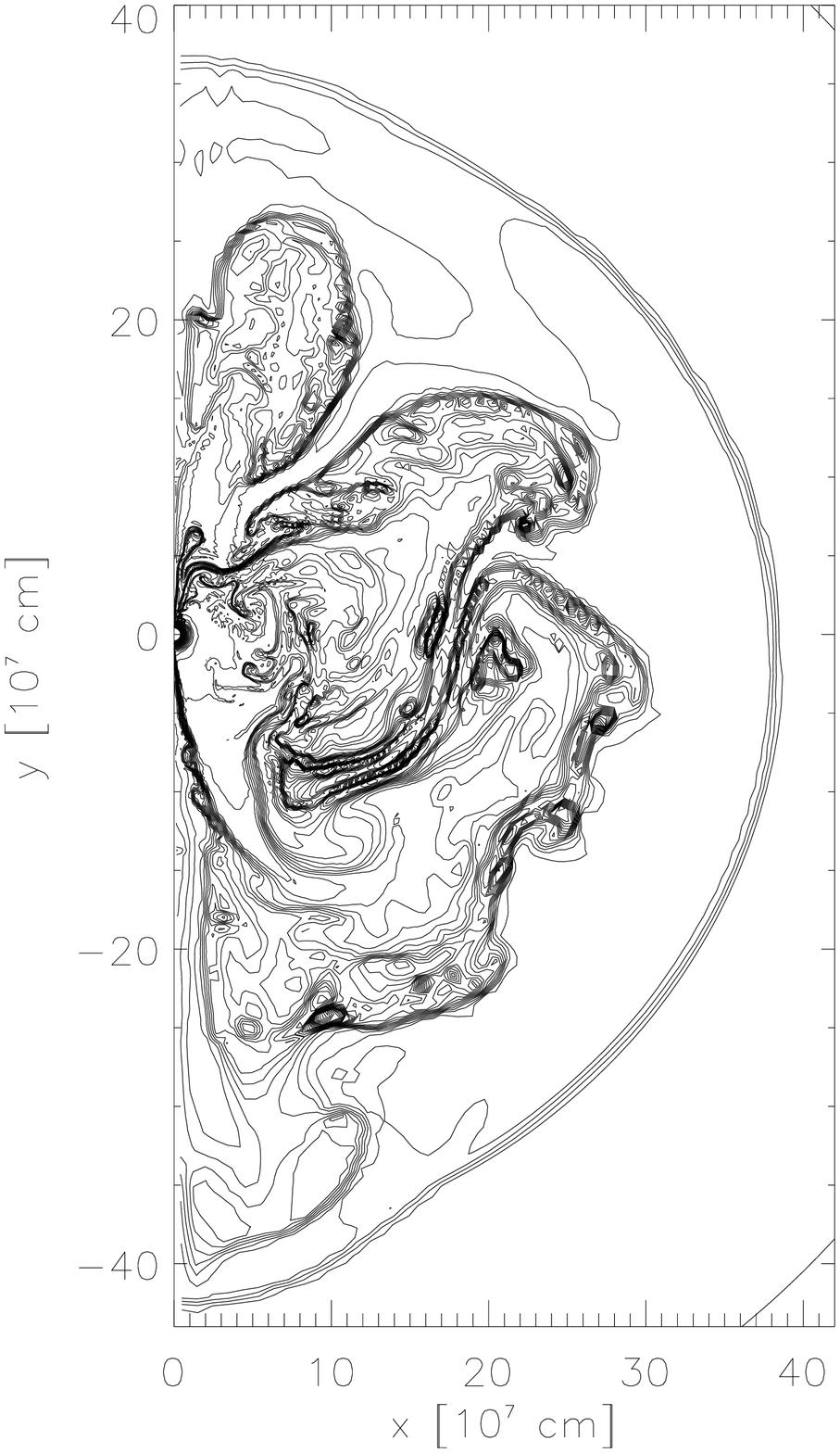}
\vskip 0.25 truecm
\par\noindent
\hskip 0.05\hsize
\vbox{
\hsize = 1.00\hsize
\baselineskip 11.0 pt
\noindent
{\bf Fig.~9.}
A weakly exploding 2D model with an explosion energy of
about $0.5\cdot 10^{51}\,{\rm erg}$
at a time $t = 377\,{\rm ms}$ after bounce.
The figure shows entropy contours, equidistantly spaced
in steps of $0.5\,k_{\rm B}/{\rm nucleon}$ between
$5\,k_{\rm B}/{\rm nucleon}$ and $16\,k_{\rm B}/{\rm nucleon}$
and in steps of $1.0\,k_{\rm B}/{\rm nucleon}$ between
$16\,k_{\rm B}/{\rm nucleon}$ and $23\,k_{\rm B}/{\rm nucleon}$.
The shock is located at about 3800~km.
Clumpy structures, inhomogeneities, and vortex patterns caused by
Kelvin-Helmholtz instabilities are visible along the boundaries
between downflows of cold gas from the postshock region
and hot, rising gas from the neutrino-heated zone around the 
protoneutron star at the coordinate center.
}
\par\noindent
\endinsert

\bigskip\noindent
{\sl 3.2.4$\ $From core bounce to one second}
\medskip\noindent
Convective overturn in the neutrino-heated layer around the 
protoneutron star develops
within $\sim 50$--$100\,{\rm ms}$ after shock formation.
About 200--$300\,{\rm ms}$ after bounce neutrinos have
deposited a sizable amount of energy in the material below the
shock front. The turbulent layer begins to
move away from the region of strongest neutrino heating and
to expand outward behind the accelerating shock (Fig.~9).
At this time turbulent activity around and close to the
protoneutron star comes to an end.
An extended phase of convection and accretion outside the
protoneutron star does not occur.
Inflows of low-entropy, $p$-rich gas
from the postshock region towards the neutrino-heated zone
are not accreted onto the neutron star. Although the
gas loses lepton number while falling in, it does not get as
$n$-rich as the material inside the gain radius. In addition,
neutrino heating and mixing with the surrounding, high-entropy
gas increase the entropy in the downflows. Both high electron
(proton) concentration and high entropy have a stabilizing
effect and prevent the penetration of the gas through the
gain radius into the cooler and more $n$-rich
surface layer of the protoneutron star.

At $\sim 400$--$500\,{\rm ms}$ the protoneutron star has become
quite compact and the density outside
has dropped appreciably. This indicates the formation of the
high-entropy, low-density ``hot-bubble" region
(Bethe \& Wilson 1985) and
the phase of small mass loss from the nascent neutron star in
the neutrino-driven wind, accompanied by slowly increasing entropies.
The wind accelerates because of the steepening density
decline away from the shrinking neutron star. The faster
expansion and push of the wind create a density
inversion between the massive, slow, inert shell behind
the shock and the evacuating hot-bubble region. Around the
time the outgoing supernova shock passes the entropy and
composition step of the Si-O interface at
$\sim 5700\,{\rm km}$ (see Fig.~8), this density inversion
steepens into a strong reverse shock that forms a sharp
discontinuity in the neutrino wind, slowing down the wind
expansion from $> 2\times 10^9\,{\rm cm/s}$ to a few
times $10^8\,{\rm cm/s}$. Since the velocities
of the wind and of the layer behind the shock
decrease with time, it is possible that this reverse shock
will trigger
fallback of a significant fraction of the matter that was
blown out in the neutrino wind. Once the
infall of the outer wind material is initiated and the pressure
support for the gas further out vanishes, inward acceleration
might even enforce the fallback of more slowly moving parts
of the dense shell behind the supernova shock.

Fallback of a significant amount of matter,
$\sim 0.1$ to $0.2\,M_\odot$, might be needed to
solve two major problems in the current supernova
models. On the one hand, due to the fast development of the explosion
and the lack of an extended accretion phase,
the protoneutron star formed at the center
of the explosion has quite a small (initial)
baryonic mass, only $\sim 1.2\,M_\odot$
in case of our $15\,M_\odot$ star with $1.3\,M_\odot$
Fe-core. On the other hand, the yields of Fe-peak elements by
explosive nucleosynthesis are incompatible with observational
constraints for type-II supernovae as
deduced from terrestrial abundances and galactic evolution arguments.
In case of powerful explosions with energies of
1--$1.3\times 10^{51}\,{\rm erg}$
material of $\sim 0.2\,M_\odot$ is heated to temperatures
$T >4.5\times 10^9\,{\rm K}$ and is ejected behind the shock during
the early phase of the explosion. Only roughly half of this matter,
0.085--$0.1\,M_\odot$,
has an electron fraction $Y_e > 0.49$ and will end up with
$^{56}$Ni as the dominant nucleosynthesis product. In that respect
the models seem to match the observations quite well.
Yet, only some part ($\sim 0.05\,M_\odot$)
of the matter that is shock-heated to $T>4.5\times 10^9\,{\rm K}$
has $Y_e \ga 0.495$ and will end up with relative
abundance yields in acceptable agreement with solar-system values.
The amount
of $^{56}$Ni produced in neutrino-driven explosions turns out to be
correlated with the explosion energy.
In case of more energetic explosions the shock
is able to heat a larger mass to sufficiently high temperatures.
\bigskip\noindent
{\sl 3.2.5$\ $Consequences}
\medskip\noindent
Turbulent overturn between the zone of strongest neutrino heating
and the supernova shock aids the re-expansion of the stalled
shock and is able to cause powerful type-II supernova explosions
in a certain, although rather narrow, window of core neutrino fluxes
where 1D models do not explode. The turbulent activity outside and
close to the protoneutron star is transient and between 300 and
$500\,{\rm ms}$ after core bounce the (essentially) spherically
symmetrical neutrino-wind phase starts and the turbulent shell
moves outward behind the expanding supernova shock. Our 2D
simulations do not show a long-lasting period of convection
and accretion after core bounce. Only very little of the cool,
low-entropy matter that flows down from the shock front to the
zone of neutrino energy deposition is advected into the
protoneutron star surface. Since the matter is $p$-rich
and its entropy increases quickly due to neutrino heating, it stays
in the heated region to gain more energy by neutrino interactions
and to start rising again.
The strong, large-scale inhomogeneities and anisotropies in the
expanding layer behind the outward propagating shock front
will probably help to explain the effects of macroscopic mixing
seen in SN~1987A and can account for neutron star recoil
velocities of a few $100\,{\rm km/s}$
(details in Janka \& M\"uller 1994, 1995b).

Although the models develop energetic explosions
for sufficiently high neutrino luminosities and produce an
amount of $^{56}$Ni that is in good agreement with observational
constraints, the initial mass of the protoneutron star is clearly
on the low side of the spectrum of measured neutron star masses.
Moreover, the models eject $\sim 0.1$--$0.15\,M_\odot$ of material
with $Y_e < 0.495$, which implies an overproduction of certain
elements in the Fe-peak by an appreciable factor compared
with the nucleosynthetic composition in the solar system.
The fallback of a significant fraction of this matter to the neutron
star at a later stage would ease these problems.
It is possible that the reverse shock which develops in our
models will trigger this fallback on a timescale of
seconds. Due to the strong inhomogeneities
in the dense layer behind the shock
this fallback could happen with considerable anisotropy and impart
an additional kick to the neutron star (Janka \& M\"uller 1995b).

On the other hand, it is hard to see how fallback could achieve
a clean disentanglement of desirable and undesirable ejecta, in 
particular if one has in mind the turbulent activity and the mixing
of different conditions present in the neutrino-heated layer.
Nevertheless,
it may still be that the observed nucleosynthetic composition of
the interstellar medium reflects the accidental result of a 
delicate separation of ``good'' and ``bad'' products of explosive
nucleosynthesis during the early phases of the explosion. 
Alternatively,
as discussed in Sect.~3.1.3, the contamination of the supernova
ejecta with overproduced $N = 50$ nuclei could be more naturally
and plausibly avoided if the luminosities of $\nu_e$ and 
$\bar\nu_e$ and therefore the degree of neutronization in the
neutrino-heated material were modified by convection in the
nascent neutron star.
\bigskip\medskip\noindent
{\bf 4.$\ $Summary and discussion}
\bigskip\smallskip\noindent
Recent multi-dimensional simulations of type-II supernova
explosions (Herant et al.~1994, Burrows et al.~1995,
Janka \& M\"uller 1996) have shown that convective overturn
in the neutrino-heating region helps the explosion and
can be crucial for the success of the delayed explosion
mechanism. The net efficiency of neutrino energy deposition
is increased when cold (low-entropy) material from behind
the shock can move inward to the region of strongest
heating, while at the same time heated (high-entropy) gas
can rise outward and expand, thus reducing the energy loss
by reemission of neutrinos. The delicate balance of
neutrino heating and cooling which is present in the
spherically symmetric case is avoided, and the neutrino luminosity
that is required for sufficiently strong heating to obtain an 
explosion is lowered.

However, there is still a competition between the 
neutrino heating timescale and the timescale for the growth of
the convective instability on the one hand and the advection
timescale of matter from the shock through the gain radius
(interior of which neutrino heating is superseded by neutrino
cooling) towards the protoneutron star on the other.
For too low neutrino luminosities the heating in the 
postshock region is not strong enough to ensure short
growth timescales of the convective overturn. In this case
the matter is advected downward through the gain radius 
faster than it is driven outward by buoyancy forces. 
Inside the gain radius the absorbed neutrino energy is 
quickly radiated away again by neutrino emission.  
Therefore convective overturn in the neutrino-heated region 
is crucial for the explosion only in a rather narrow window
of luminosities. 
For higher core neutrino fluxes also spherically symmetrical
models yield energetic explosions, while for lower luminosities
even with convection no strong explosions occur.
In any case, the success of the delayed explosion mechanism
requires sufficiently large neutrino luminosities from the
nascent neutron star for a sufficiently long time.

Moreover, all current supernova models have severe problems
concerning the nucleosynthetic composition of the neutrino-heated
ejecta. These show huge overproduction factors (up to the order of
100) for elements around neutron number $N=50$, indicating 
that the ejecta are too neutron-rich. Several suggestions have
been made to solve this problem, e.g., fallback of some of
the early ejecta towards the neutron star during a later phase
of the explosion (Herant et al.~1994), a longer delay of
the explosion than obtained in current models which would reduce
the amount of neutron-rich material due to the density decrease in
the region between supernova shock and protoneutron star
(Burrows \& Hayes 1995), or a slightly 
underestimated electron fraction $Y_e$ in the ejecta because
of still unsatisfactorily treated neutrino transport which 
affects the computed $\nu_e$ and $\bar\nu_e$ spectra and thus
the neutrino-matter interactions in the hot bubble region
(Hoffman et al.~1996). While the latter might certainly be
true and has to be investigated carefully (see Sect.~2.1), 
all of these 
suggestions rely on some fine-tuning and effects which might
be very unstable and sensitive to minor changes. 

Strong convection inside the newly formed neutron star 
during the first few hundred milliseconds after core bounce
and shock formation offers a remedy of both problems
(Keil et al.~1996).
Faster cooling of the protoneutron star by convective 
energy transport increases the total neutrino luminosities 
and therefore helps the neutrino energy deposition in the postshock
layers. The accompanying enhanced deleptonization raises the
$\nu_e$ number flux relative to the $\bar\nu_e$ number flux.
This leads to more frequent absorptions of $\nu_e$ onto
neutrons in the neutrino-heated gas and thus can establish
a higher electron fraction in the material that carries the
supernova energy during the early moments of the explosion.
The final confirmation of this picture has still to be waited
for until multi-dimensional supernova simulations 
have been performed that follow
the protoneutron star {\it and} the surrounding progenitor star
for a sufficiently long time.

%
%
\bigskip\noindent
{\bf Acknowledgements.}$\ $ 
The author would like to thank the organizers, 
in particular M.~Locher, for the great hospitality and the 
opportunity to attend the summer school in Zuoz. 
This work was supported
by the Sonderforschungsbereich 375-95 for 
Astro-Particle Physics of the Deutsche Forschungsgemeinschaft.
The computations were performed on the CRAY-YMP 4/64 of the
Rechenzentrum Garching.
%

%
%
\bigskip\medskip
{
\petit{
\baselineskip 10pt
\parskip = 3pt plus 1pt minus 1pt
\medskip\noindent
{\bf References}
\medskip\smallskip
\noindent
\refs

Alexeyev E.N., et al., 1988, Phys.~Lett.~B205, 209

Arnett W.D., 1988, ApJ 331, 337

Arnett W.D., Bahcall J.N., Kirshner R.P., Woosley S.E., 1989,
ARA\&A 27, 629

Arnett W.D., Fryxell B.A., M\"uller E., 1989, ApJ 341, L63

Aschenbach B., Egger R., Tr\"umper J., 1995, Nat 373, 587

Barthelmy S., 1989, et al., IAU Circ.~4764

Bethe H.A., Wilson J.R., 1985, ApJ 295, 14

Bethe H.A., Brown G.E., Cooperstein J., 1987, ApJ 322, 201
 
Bionta R.M., et al., 1987, Phys.~Rev.~Lett.~58, 1494 (IMB collaboration)

Bruenn S.W., 1985, ApJS 58, 771

Bruenn S.W., 1993, in: Nuclear Physics in the Universe,
eds.~M.W.~Guidry and M.R.~Strayer, IOP, Bristol, p.~31

Bruenn S.W., Dineva T., 1996, ApJ~458, L71

Bruenn S.W., Mezzacappa A., 1994, ApJ 433, L45

Bruenn S.W., Mezzacappa A., Dineva T., 1995, Phys.~Rep.~256, 69

Burrows A., 1987, ApJ 318, L57

Burrows A., Fryxell B.A., 1992, Sci 258, 430

Burrows A., Fryxell B.A., 1993, ApJ 418, L33

Burrows A., Goshy J., 1993, ApJ 416, L75

Burrows A., Hayes J., 1995, Ann.~N.Y.~Acad.~Sci.~759, 375

Burrows A., Hayes J., Fryxell B.A., 1995, ApJ~450, 830

Burrows A., Lattimer J.M., 1986, ApJ 307, 178
 
Burrows A., Lattimer J.M., 1988, Phys.~Rep.~163, 51

Caraveo P.A., 1993, ApJ 415, L111
 
Colella P., Woodward P.R., 1984, J.~Comp.~Phys.~54, 174

Colgan S.W.J., Haas M.R., Erickson E.F., Lord S.D., Hollenbach D.J.,
1994, ApJ 427, 874

Colgate S.A., Herant M., Benz W., 1993, Phys.~Rep.~227, 157

Colgate S.A., White R.H., 1966, ApJ 143, 626

Cook W.R., et al., 1988, ApJ 334, L87

Den M., Yoshida T., Yamada Y., 1990, Prog.~Theor.~Phys.~83, 723

Dotani T., et al., 1987, Nat 330, 230

Erickson E.F., et al., 1988, ApJ 330, L39

Epstein R.I., 1979, MNRAS 188, 305

Frail D.A., Kulkarni S.R., 1991, Nat 352, 785

Fryxell B.A., M\"uller E., Arnett W.D., 1989, MPA-Preprint 449,
Max-Planck-Institut f\"ur Astrophysik, Garching

Fryxell B.A., M\"uller E., Arnett W.D., 1991, ApJ 367, 619

Gehrels N., Leventhal M., MacCallum C.J., 1988, in: Nuclear Spectroscopy
of Astrophysical Sources, eds.~N.~Gehrels and G.~Share,
AIP, New York, p.~87

Guidry M.W., 1996, talk at the workshop of the `European
Centre for Theoretical Studies in Nuclear Physics'
on Physics of Su\-per\-no\-vae and Neutron Stars,
Trento, Italy, June 3--14, 1996

Haas M.R., et al., 1990, ApJ 360, 257

Hachisu I., Matsuda T., Nomoto K., Shigeyama T., 1990, ApJ 358, L57

Hachisu I., Matsuda T., Nomoto K., Shigeyama T., 1991, ApJ 368, L27

Harrison P.A., Lyne A.G., Anderson B., 1993, MNRAS 261, 113

Herant M., Benz W., 1991, ApJ 345, L412

Herant M., Benz W., 1992, ApJ 387, 294

Herant M., Benz W., Colgate S.A., 1992, ApJ 395, 642

Herant M., Benz W., Hix W.R., Fryer C.L., Colgate S.A., 1994,
ApJ 435, 339

Hillebrandt W., 1987, in: High Energy Phenomena Around Collapsed Stars,
NATO ASI~C195,
ed.~F.~Pacini, Reidel, Dodrecht, p.~73
 
Hirata K., et al., 1987,
Phys.~Rev.~Lett.~58, 1490 (Kamiokande II collaboration)

Hoffman R.D., Woosley S.E., Fuller G.M., Meyer B.S.,
1996, ApJ~460, 478

Janka H.-Th., 1993, in:
Frontier Objects in Astrophysics and Particle Physics,
eds.~F.~Giovannelli and G.~Mannocchi,
Societ\`a Italiana di Fisica, Bologna, p.~345

Janka H.-Th., 1995, in:
Proc.~of the Workshop on Astro-Particle Physics of the
Sonderforschungsbereich 375,
Ringberg Castle, Tegernsee, March 6--10, 1995,
eds.~A.~Weiss, G.~Raffelt, W.~Hillebrandt, F.~von Feilitzsch,
Technische Universit\"at M\"unchen, p.~154 

Janka H.-Th., M\"uller E., 1993, in: Frontiers of Neutrino
Astrophysics, eds.~Y.~Suzuki and K.~Nakamura,
Universal Academy Press, Tokyo, p.~203

Janka H.-Th., M\"uller E., 1994, A\&A 290, 496

 
Janka H.-Th., M\"uller E., 1995a, ApJ~448, L109 

Janka H.-Th., M\"uller E., 1995b, Ann.~N.Y.~Acad.~Sci.~759, 269

Janka H.-Th., M\"uller E., 1996, A\&A~306, 167

Janka H.-Th., Keil W., Raffelt G., Seckel D., 1996, 
Phys.~Rev.~Lett.~76, 2621

Keil W., 1996, Ph.D.~Thesis, TU M\"unchen (in preparation)

Keil W., Janka H.-Th., 1995, A\&A~296, 145

Keil W., Janka H.-Th., M\"uller E., 1996, 
MPA-preprint 971, ApJL, in press
 
Keil W., Janka H.-Th., Raffelt G., 1995, Phys.~Rev.~D 51, 6635

Lattimer J.M., Swesty F.D., 1991, Nucl.~Phys.~A535, 331

Li H., McCray R., Sunyayev R.A., 1993, ApJ 419, 824

Lyne A.G., Lorimer D.R., 1994, Nat 369, 127

Mahoney W.A., et al., 1988, ApJ 334, L81

Matz S.M., et al., 1988, Nat 331, 416

Mayle R.W., Wilson J.R., 1988, ApJ~334, 909

McLaughlin G.C., Fuller G.M., Wilson J.R., 1996, ApJ, in press

Miller D.S., Wilson J.R., Mayle R.W., 1993, ApJ 415, 278

M\"uller E., 1993, in: Proc.~of the 7th Workshop on Nuclear
Astrophysics (Ringberg Castle, March 22-27, 1993),
eds.~W.~Hillebrandt and E.~M\"uller, Report MPA/P7, Max-Planck-Institut
f\"ur Astrophysik, Garching, p.~27

M\"uller E., Fryxell B.A., Arnett W.D., 1991, in: ESO/EIPC Workshop
on SN~1987A and other Supernovae, ESO Workshop and Conference
Proceedings No.~37,
eds.~I.J.~Danziger and K.~Kj\"ar, ESO, Garching, p.~99

M\"uller E., Janka H.-Th., 1994, in: Reviews in Modern Astronomy 7,
Proceedings of the International Scientific Conference of the AG
(Bochum, Germany, 1993), ed.~G.~Klare,
Astronomische Gesellschaft, Hamburg, p.~103

Qian Y.-Z., Woosley S.E., 1996, ApJ, in press

Raffelt G., Seckel D., 1991, Phys.~Rev.~Lett.~67, 2605

Rank D.M., 1988, et al., Nat 331, 505

Sandie W.G., et al., 1988, ApJ 334, L91

Shigeyama T., Nomoto K., 1990, ApJ 360, 242

Shigeyama T., Nomoto K., Hashimoto M., 1988, A\&A 196, 141

Shimizu T., Yamada S., Sato K., 1993, PASJ 45, L53

Shimizu T., Yamada S., Sato K., 1994, ApJ 432, L119

Sigl G., 1996, Phys.~Rev.~Lett.~76, 2625

Spyromilio J., Meikle W.P.S., Allen D.A., 1990, MNRAS 242, 669

Stewart R.T., Caswell J.L., Haynes R.F., Nelson G.J., 1993,
MNRAS 261, 593

Sumiyoshi K., Suzuki H., Toki H., 1995, A\&A~303, 475

Sunyaev R.A., et al., 1987, Nat 330, 227
 
Suzuki H., 1989, Ph.D.~Thesis, Univ.~of Tokyo

Takahashi K., Witti J., Janka H.-Th., 1994, A\&A 286, 857

Taylor J.H., Manchester R.N., Lyne A.G., 1993, ApJS 88, 529

Teegarden B.J., et al., 1989, Nat 339, 122

Tueller J., et al., 1990, ApJ 351, L41

Wilson J.R., Mayle R.W., 1988, Phys.~Rep.~163, 63

Wilson J.R., Mayle R.W., 1989, in: The Nuclear
Equation of State, Part~A, eds.~W.~Greiner and
H.~St\"ocker, Plenum Press, New York, p.~731

Wilson J.R., R.W.~Mayle, 1993, Phys.~Rep.~227, 97

Wilson R.B., et al., 1988, in: Nuclear Spectroscopy of Astrophysical Sources,
eds.~N.~Gehrels and G.~Share, AIP, New York, p.~66

Witteborn F., et al., 1989, ApJ 338, L9

Witti J., Janka H.-Th., Takahashi K., 1994, A\&A 286, 841

Woosley S.E., 1988, ApJ 330, 218

Woosley S.E., Hoffman R.D., 1992, ApJ~395, 202

Woosley S.E., Pinto P.A., Ensman L., 1988, ApJ 324, 466

Woosley S.E., Wilson J.R., Mathews G.J., Hoffman R.D., Meyer B.S.,
1994, ApJ 433, 229

Yamada Y., Nakamura T., Oohara K., 1990, Prog.~Theor.~Phys.~84, 436

Yamada S., Shimizu T., Sato K., 1993, Prog.~Theor.~Phys.~89, 1175

\endrefs
}
}

\vfill\eject\end\bye